\documentclass[twocolumn,floatfix,pra,aps,showpacs]{revtex4}
\usepackage{epsfig}
\usepackage{graphicx}
\usepackage{amsmath}
\usepackage{color}

\newcommand{\be}{\begin{equation}}
\newcommand{\ee}{\end{equation}}
\newcommand{\mb}{\mathbf}
\newcommand{\mc}{\mathcal}

\begin{document}

\title{Persistent currents in coherently coupled Bose-Einstein condensates in a ring trap}
\author{Marta Abad}
\affiliation{Quantum Systems Unit, Okinawa Institute of Science and Technology Graduate University, Okinawa, 904-0495, Japan}
\date{\today}

\begin{abstract}
We study the stability of persistent currents in a coherently coupled quasi-2D Bose-Einstein condensate confined in a ring trap at $T=0$. By numerically solving Gross-Pitaevskii equations and by analyzing the excitation spectrum obtained from diagonalization of the Bogoliubov-de Gennes matrix, we describe the mechanisms responsible for the decay of the persistent currents depending on the values of the interaction coupling constants and the Rabi frequency. When the unpolarized system decays due to an energetic instability in the density channel, the spectrum may develop a roton-like minimum, which gives rise to the finite wavelength excitation necessary for vortex nucleation at the inner surface. When decay in the unpolarized system is driven by spin-density excitations, the finite wavelength naturally arises from the existence of a gap in the excitation spectrum. In the polarized phase of the coherently coupled condensate, there is an hybridization of the excitation modes that leads to complex decay dynamics. In particular, close to the phase transition, a state of broken rotational symmetry is found to be stationary and stable.
\end{abstract}

\maketitle

\section{Introduction}

Persistent currents have been one of landmarks of quantum fluids since their discovery. The emergence of a complex order parameter below the superfluid phase transition leads to the quantization of the circulation of the superfluid velocity. In superconductors, since it is charged particles that move, this property is related to the quantization of magnetic flux and has led, among other applications, to the creation of superconducting quantum interference devices (SQUIDs), which are present nowadays in many laboratories. Helium superfluids behave in an analogous way, but it is now a neutral current of atoms that flows, allowing for instance for the Hess-Fairbaink effect (equivalent to the Meissner effect in superconductors), or for proposals of precise rotation sensors (equivalent to SQUIDs in superconductors).

Bose-Einstein condensates (BECs) offer a unique framework to study persistent currents, since the absence of electric charge and normal components (if cooled down well below the critical temperature) allows one to isolate the effects arising from the purely quantum nature of the fluid. From a theoretical point of view, the decay of persistent currents is linked to the energetic Landau instability of a moving superfluid, which in a ring geometry is closely related to Bloch's argument~\cite{Bloch} for the disappearance of metastable states. In a different picture, this metastability of the persistent currents can be linked to the presence of energy barriers between states with different angular momentum~\cite{Benakli1999,Tempere2001,Capuzzi2009}. Persistent currents in BECs have been observed experimentally~\cite{Ryu2007,Moulder2012,Kumar2015,Murray2013}, and their decay has been seen to be driven by a combination of flow instabilities and stochastic events~\cite{Moulder2012}. 
This physics is slightly different from the decay of the superfluid flow through a weak link, where the instabilities are triggered within the barrier region, and is the subject of extensive theoretical and experimental investigation~\cite{Piazza2009,Watanabe2009,Ramanathan2011,Eckel2014X,Finazzi2015,Yakimenko2015,Munoz2015}. Analogously to superconductors and helium superfluids, research is advancing towards the creation of atomic SQUID-like interferometers~\cite{Wright2013,Edwards2013,Ryu2013,Eckel2014,Jendrzejewski2014}, which due to the diversity of interactions in BECs could even have a self-induced character~\cite{Abad2011}.
 
We consider persistent currents in a special kind of BEC: a coherently coupled two-component condensate, where population transfer between two hyperfine levels is allowed via Raman transitions. This system is at the basis of more complex systems such as spin-orbit coupled BECs, and presents unique features (for a review, see \cite{Abad2013} and references therein). In particular, the ground state shows a magnetic-like transition between neutral and polarized states and the excitation spectrum in free space is characterized by two modes: a gapless sound-like mode and a gapped mode that show density or spin-density characters or a combination of both (hybridization). These features play a very important role in the stability and decay of persistent currents, which differ substantially from binary mixtures \cite{Smyrnakis2009,Beattie2013,Anoshkin2013,Yakimenko2013,Abad2014,Wu2015} and can help understand superfluidity in systems characterized by more complex order parameters.

Coherently coupled two-component condensates have received strong attention. In particular, several superfluid properties have been addressed, such as: internal Josephson effect~\cite{Zibold2010}; split vortices~\cite{GarciaRipoll2002} and domain walls~\cite{Son2002}; vortex lattices~\cite{Cipriani2013}; phase winding in the presence of Rabi coupling \cite{Hamner2013}; soliton~\cite{Usui2015} and vortex \cite{Gallemi2015} dynamics arising from counterflow; dipole oscillations \cite{Sartori2015}; dynamics following a quench~\cite{Nicklas2015}; and vortices and persistent currents in state-selective potentials of different geometries~\cite{Ino2015}. 

In this work we concentrate on the stability and decay of stationary persistent currents when both components are trapped in a ring-shaped potential. We characterize the stability conditions and decay dynamics by solving the Gross-Pitaevskii equation and analyzing the linearized excitation spectrum given by the fully numerical Bogoliubov-de Gennes equations. We study the decay mechanisms both in the neutral and the polarized phases, and understand them in terms of energetically unstable modes propagating along the azimuthal direction. We discuss the difference in behavior when decay is induced by pure density or pure spin excitations, showing that in the former case the spectrum acquires a roton-like structure that allows the formation of the vortices responsible for phase-slippage, while in the latter case the vortices are formed directly at the minimum of the Doppler-shifted spectrum. In the polarized phase we go from a situation far from the phase transition where excitations assume a single-component character, to a situation close to the phase transition where there is a breaking of the rotational symmetry and the system develops striking density structures, which are stable and stationary configurations. 

The article is organized as follows. In Sec.~\ref{SecTheory} we describe the theoretical framework in which the work is carried out, namely mean-field Gross-Pitaevskii equation and Bogoliubov-de Gennes equations for linear perturbations. In Sec.~\ref{SecGS} we describe the ground state of the coherently coupled BEC in a ring trap. In Secs.~\ref{SecGS1} and \ref{SecGS2} the stability of persistent currents is studied, respectively, in the neutral and polarized ground states. Finally, further discussions and the conclusions are drawn in Sec.~\ref{SecConcl}.

\section{Theoretical framework}\label{SecTheory}

\subsection{Gross-Pitaevskii equation}\label{SecTheoryGP}

We consider a $T=0$ Bose-Einstein condensate that consists of two hyperfine states coherently coupled by a Rabi frequency, in the regime of tight axial harmonic trapping, where the system can be effectively considered two-dimensional (2D).
Within mean-field regime this system is described (in the rotating wave approximation) by the coupled Gross-Pitaevskii equations (GP)
\begin{align}
	i\hbar\frac{\partial \Psi_1}{\partial t} =& \left[ -\frac{\hbar^2}{2m}\nabla^2 + V_1 + g_{11}|\Psi_1|^2 + g_{12}|\Psi_2|^2 \right]\Psi_1 +\nonumber\\
	 &+  \Omega_R \Psi_2 \label{Eq:GP1}\\
	i\hbar\frac{\partial \Psi_2}{\partial t} =& \left[ -\frac{\hbar^2}{2m}\nabla^2 + V_2 + g_{22}|\Psi_2|^2 + g_{12}|\Psi_1|^2 \right]\Psi_2 +\nonumber\\
	 &+  \Omega_R \Psi_1 \label{Eq:GP2}
\end{align}
where $\Psi_\alpha$ are the wave functions of components $\alpha=1,2$, $V_\alpha\equiv V$ are the external trapping potentials, $g_{\alpha\alpha}$ and $g_{12}$ are respectively the intra- and inter-species coupling constants, and $2\Omega_R/\hbar$ is the generalized Rabi frequency given by the interaction of the atoms with the laser fields. In this setup $\Omega_R$ is positive and real~\cite{comment3}, uniform in space and constant in time, thus not leading to any spin-orbit couplings. 
After integration along the axial direction, the interaction constants relate to the scattering lengths as $g_{\gamma} = \sqrt{8\pi\lambda} a_\gamma\hbar\omega_\perp a_\perp$ ($\gamma=11, 22, 12$), with $a_\gamma$ the 3D s-wave scattering length, $\lambda=\omega_z/\omega_\perp$ the trap asymmetry, $a_\perp=\sqrt{\hbar/m\omega_\perp}$ the radial oscillator length, and $\omega_\perp$ and $\omega_z$ the radial and axial harmonic trap frequencies.
The ring potential is simulated as a harmonic plus Gaussian potential, 
\be
	V = \frac{1}{2}m\omega_\perp^2 r_\perp^2 + V_0 e^{-2r_\perp^2/\sigma_0^2}
\ee
where $r_\perp=x^2 + y^2$, and $V_0$ and $\sigma_0$ are given by the laser intensity and the beam waist of the laser digging the hole. For concreteness, in the simulations we have taken the case of $^{87}$Rb atoms and have fixed the following values of the parameters: $\omega_\perp=200\times2\pi$~s$^{-1}$, $\lambda=100$, $V_0=200\,\hbar\omega_\perp$, $\sigma_0=1\,a_\perp$, $g_{11}=g_{22}=3.48\times10^{-3}\,\hbar\omega_\perp a_\perp^2$, and a total number of particles $N=10^5$. Different values of $g_{12}$ and $\Omega_R$ have been used to explore the different regimes of the system. The results are however generalizable to other atomic species, number of atoms in the condensate, laser parameters and interaction coupling constants.

This system has spinor character, and the order parameter in the superfluid phase is given by $\Psi = (\Psi_1,\Psi_2)^T$. There is only one broken U(1) symmetry, which in particular means that the stationary states take the form
\be
	\left(\begin{array}{c}
		\Psi_1(\mb{r},t)\\
		\Psi_2(\mb{r},t)
	\end{array}\right) =
	\left(\begin{array}{c}
		\psi_1(\mb{r})\\
		\psi_2(\mb{r})
	\end{array}\right) e^{-i\mu t/\hbar}	
\ee
where $\mu$ is the chemical potential.
This is in contrast to binary mixtures (where $\Omega_R=0$), where there are two coexisting condensates, that is two broken U(1) symmetries, and one can define a chemical potential for each component since the numbers of particles are conserved separately. This leads to a huge conceptual difference between coherently coupled BECs and binary mixtures \cite{Abad2013}. In particular the presence of the linear coupling $\Omega_R$ changes the order of the phase transition from first order (demixing instability in binary mixtures) to second order (ferromagnetic-like transition in the coherently coupled case), leading to a finite polarization instead of phase separation. This is seen to be the case both in the mean-field regime (homogeneous and trapped) and in the purely quantum regime of strong correlations in optical lattices \cite{Zhan2014,Barbiero2014}. 
 
The numerical simulations presented here are performed in a grid of $256\times256$ points, with a spacing of $0.06\,a_\perp$ in both directions (in some particular cases, we have also run the simulations in a grid of $512\times512$ and spacing of $0.04\,a_\perp$ to check the accuracy of the original grid, finding very good agreement). Equations~\eqref{Eq:GP1} and \eqref{Eq:GP2} are numerically solved in imaginary time to find the metastable solutions for different values of the winding number, $\kappa$, characterizing the persistent currents. This is done without any constraint during evolution, but imprinting the desired value of $\kappa$ in the initial trial wave function. The real-time dynamics are performed using a Hamming's algorithm (predictor, corrector, modifier) initialized by a 4th order Runge-Kutta method, checking that both the energy and number of particles are conserved throughout the simulation. In order to study dissipative dynamics (necessary to simulate energetic instabilities) we have added a small real term in the left-hand-side of Eqs.~\eqref{Eq:GP1} and \eqref{Eq:GP2}, that is $i\partial_t\to (i-\gamma)\partial_t$, with $\gamma=0.03$. This term (which can be thought of as a small imaginary time component) simulates a coupling to a reservoir (for instance, the thermal cloud) that can remove energy from the system. To isolate the effects of dissipation, the wave function is renormalized at each time step. 

\subsection{Bogoliubov-de Gennes excitations}\label{SecTheoryBdG}

The linear stability analysis of the stationary solutions to Eqs.~\eqref{Eq:GP1} and \eqref{Eq:GP2} is done by applying the Bogoliubov prescription to the time-dependent wave functions $\Psi_\alpha$, that is
\be
	\Psi_\alpha(\mb{r},t) = \left[\psi_\alpha(\mb{r}) + \delta\Psi_\alpha(\mb{r},t)\right]e^{-i\mu t/\hbar}\ ,
\ee
with $\psi_\alpha$ the converged wave functions obtained from imaginary time propagation of Eqs.~\eqref{Eq:GP1} and \eqref{Eq:GP2}, and $\alpha=1,2$.
Substituting these ans\"atze (and the corresponding complex conjugates) into Eqs.~\eqref{Eq:GP1} and \eqref{Eq:GP2} and keeping terms up to first order in $\delta\Psi_\alpha$ we find a set of equations, usually referred to as Bogoliubov-de Gennes (BdG), that can be written in matrix form as
\begin{equation}
	i\hbar\frac{\partial}{\partial t}\left( \begin{array}{c} \delta\Psi_1 \\ \delta\Psi_1^* \\ \delta\Psi_2 \\ \delta\Psi_2^* \end{array} \right) = \mathcal{L}\left( \begin{array}{c} \delta\Psi_1 \\ \delta\Psi_1^* \\ \delta\Psi_2 \\ \delta\Psi_2^* \end{array} \right) \label{EqBdG}
\end{equation}
where the differential operator $\mathcal{L}$ is given by
\begin{widetext}
\begin{equation}
    \mathcal{L} = \left( \begin{array}{cccc} 
    				h_1 & g_{11}(\psi_1)^2 & g_{12}\psi_1\psi_2^* + \Omega_R & g_{12}\psi_1\psi_2 \\
				-g_{11}(\psi_1^*)^2 & -h_1 &  -g_{12}\psi_1^*\psi_2^* & -g_{12}\psi_1^*\psi_2  - \Omega_R\\
				g_{12}\psi_1^*\psi_2 + \Omega_R& g_{12}\psi_1\psi_2 & h_2 & g_{22}(\psi_2)^2  \\
				-g_{12}\psi_1^*\psi_2^*  & -g_{12}\psi_1\psi_2^* - \Omega_R  &- g_{22}(\psi_2^*)^2 & -h_2 
    			 \end{array} \right) 
\end{equation}
\end{widetext}
with the diagonal terms given by
\begin{align}
	h_1 =  -\frac{\hbar^2}{2m}\nabla^2 + V_1 + 2g_{11}n_1 + g_{12}n_2 - \mu \\
	h_2 =  -\frac{\hbar^2}{2m}\nabla^2 + V_2 + 2g_{22}n_2 + g_{12}n_1 - \mu \,.
\end{align}
The operator $\mathcal{L}$ is a space-dependent, non-linear, non-hermitian operator with well-known properties (see for instance~\cite{Castin}).

In homogeneous space, the perturbations $\delta\Psi_\alpha$ can be expanded in a plane-wave basis, $\sim e^{i(\mb{k}\cdot\mb{r}-\omega t)}$, and one can find analytical solutions to the eigenvalue problem~\cite{Goldstein1997,Search2001,Tommasini2003,Abad2013}.

In the trapped system the solutions are in general not analytic and one has to solve the eigenvalue problem with numerical methods. In the most general case, the perturbations are expanded as
\be
	\left( \begin{array}{c} \delta\Psi_1(\mb{r},t) \\ \delta\Psi_1^*(\mb{r},t) \\ \delta\Psi_2(\mb{r},t) \\ \delta\Psi_2^*(\mb{r},t) \end{array} \right) \sim \left( \begin{array}{c} \mc{U}_1(\mb{r}) \\ \mc{V}_1(\mb{r}) \\ \mc{U}_2(\mb{r}) \\ \mc{V}_2(\mb{r}) \end{array} \right)e^{-i\omega t}\label{EqAnsatzUV}
\ee
and diagonalization of $\mathcal{L}$ yields a set of eigenvectors $(\mc{U}_1,\mc{V}_1,\mc{U}_2,\mc{V}_2)^T$ and corresponding eigenvalues $\omega$. Numerically this can be done by writing the differential term $\nabla^2$ appearing in $\mc{L}$ using a second-order finite-difference approximation and writing the matrix $\mc{L}$ in a spatial basis. For a calculation grid of $256\times 256$ points in $x$ and $y$ directions, the dimensions of the (sparse) matrix $\mc{L}$ are $2^{18}\times2^{18}$. We use the Lanczos method to diagonalize such a matrix, which is implemented in the function {\it eigs} of Matlab. 

After diagonalization, the eigenvectors are normalized as
\be
	\int d\mb{r} \left( |\mc{U}_1|^2 - |\mc{V}_1|^2 + |\mc{U}_2|^2 - |\mc{V}_2|^2\right) = \pm 1\ .
\ee
Positive-norm eigenvectors correspond (in our notation) to the physical solutions of the problem, while negative norm eigenvectors will be discarded. With this choice, negative frequencies indicate energetic instabilities of the system, and imaginary or complex frequencies represent dynamical instabilities (see, for instance, \cite{Castin}). Notice that for purely imaginary eigenvalues, which indicate dynamical instability, the normalization above is zero. In addition, there is a mode with eigenvalue $\omega=0$ and zero norm that represents the gauge mode, which can be understood as a global phase excitation of the reference state $\psi$, but that has no physical effect. For completeness, we will show this mode in all the spectra.

The eigenvector components $\mathcal{U}_\alpha$ and $\mathcal{V}_\alpha$ give important information on the nature of the excitations. In the absence of phase gradients in the reference state and for stable modes, the eigenvectors are real. When currents are present with winding number $\kappa$, instead, the eigenvectors become complex. In rotationally symmetric systems, they can be characterized by the azimuthal quantum number, $\ell$. The phase of the eigenvector components then shows vortex-like patterns with winding number $\ell_{\mathcal{U}}=\ell+\kappa$, $\ell_{\mathcal{V}}=\ell-\kappa$, where $\ell_{\mathcal{U}}$ and $\ell_{\mathcal{V}}$ indicate the winding numbers exhibited by the components $\mathcal{U}$ and $\mathcal{V}$ of a particular eigenvector. In addition, the eigenvectors may show radial nodes with quantum number $n_\perp$. In this work we focus on the modes with $n_\perp=0$, which are the lowest in energy and the ones driving the decay.

The dynamical evolution of $\Psi_\alpha$ is a result of a combination of Bogoliubov modes. However, to understand the effect of a particular eigenmode, it is convenient to write the perturbation that it creates on the wave function as
\be
	\Psi_\alpha(\mathbf{r},t)\sim\psi_\alpha(\mathbf{r})+\varepsilon\left[\mathcal{U}_\alpha(\mathbf{r})e^{-i\omega t} + \mathcal{V}_\alpha^*(\mathbf{r})e^{i\omega t}\right]\label{EqPerturb}
\ee
with $\alpha=1,2$ and $\varepsilon$ a number quantifying the proportion of perturbation added to the reference state wave function ($\varepsilon\ll 1$ to remain in the Bogoliubov regime).  
This representation of the modes is very convenient to understand the change in the condensate wave function due to a particular excitation. 
The above expression, moreover, can be used as initial wave function in the dynamics of Eqs.~\eqref{Eq:GP1} and \eqref{Eq:GP2} to facilitate the development of excitations.

\section{Ground state structure}\label{SecGS}

Following the notation in Ref.~\cite{Abad2013}, when $g_{11}=g_{22}\equiv g$, the ground state of the system is characterized by the presence of a neutral (GS1) and a polarized (GS2) configurations. The phase transition between the two is of the second order and in the homogeneous system it takes place when the equality $g_{12}=g+2\Omega_R/n$ is satisfied, with $n$ the total density, $n_1=n_2\equiv n/2$ . This allows us to introduce a critical value for one of the parameters $g_{12}$, $g$ or $\Omega_R$. In this work we will choose the latter one, therefore we will have
\be
	\Omega_c= \frac{n}{2}(g_{12}-g)\,.\label{EqOmegac}
\ee
The system is in GS1 when $\Omega_R\ge\Omega_c$, and in GS2 otherwise.
The condition $g_{11}\neq g_{22}$ has the effect of creating a permanent polarization in the condensate and the behavior of the system in a way resembles that in GS2. For the sake of simplicity and clarity, we will restrict here to the case where the intra-component interaction coupling constants are equal.

\begin{figure}\centering
	\includegraphics[clip=true,width=\linewidth]{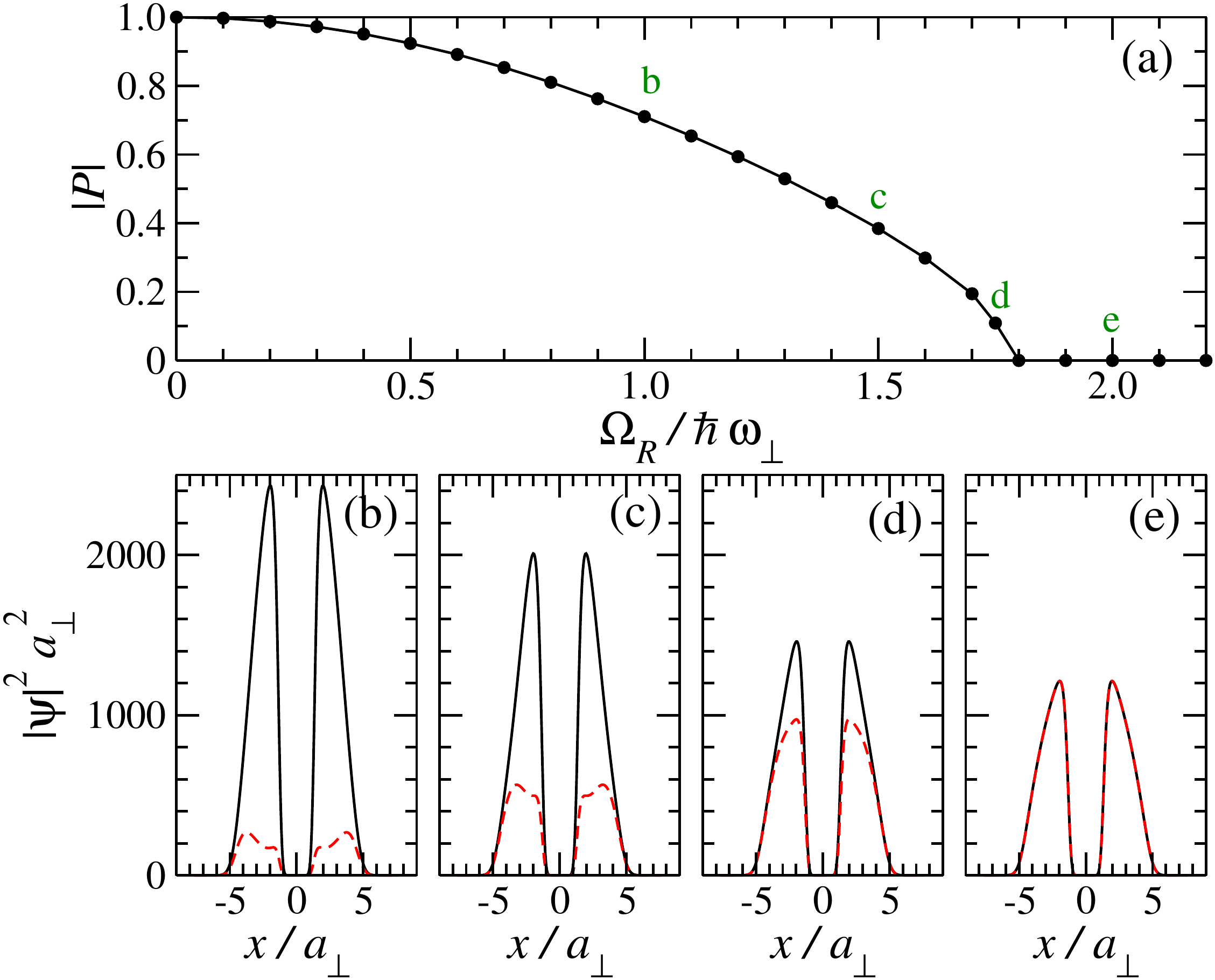}
	\caption{(a) Polarization phase transition in the ring trap, for $g_{12}=1.5g$. Bottom panels: Examples of the density profiles along the $x$ direction through the phase transition: (b) $\Omega_R=1\hbar\omega_\perp$, (c) $\Omega_R=1.5\hbar\omega_\perp$, (d) $\Omega_R=1.75\hbar\omega_\perp$, (e) $\Omega_R=2\hbar\omega_\perp$. Majority and minority components are plotted as solid and dashed lines, respectively.}\label{Fig1}
\end{figure}

In the presence of an external trap the density depends on position, $n({\bf r})$, and the equality above is reached for some value $r = R_{c}$. This means that in a confined system if $\Omega_R<\Omega_c$ the two phases GS1 and GS2 coexist, with the GS1 phase at the surface of the condensate \cite{Abad2013}. 
The global GS1--GS2 transition can still be characterized by the polarization of the system, $P=(N_1-N_2)/N$, where $N_1$ and $N_2$ are the numbers of particles in components $1$ and $2$, respectively. This quantity is shown in the top panel of Fig.~\ref{Fig1} as a function of $\Omega_R$ for the ring trap. Examples of ground state densities for different values of $\Omega_R$ can be seen in the bottom panels of Fig.~\ref{Fig1}. We can clearly see that in the ring trap there exist two critical radii $R_c^{\pm}$ where the transition from GS1 to GS2 takes place. 
Notice that in the ground state there is always a $\pi$ phase-shift between the phases of the wave functions for components $1$ and $2$, since we take $\Omega_R>0$.

\begin{figure}\centering
	\includegraphics[clip=true,width=\linewidth]{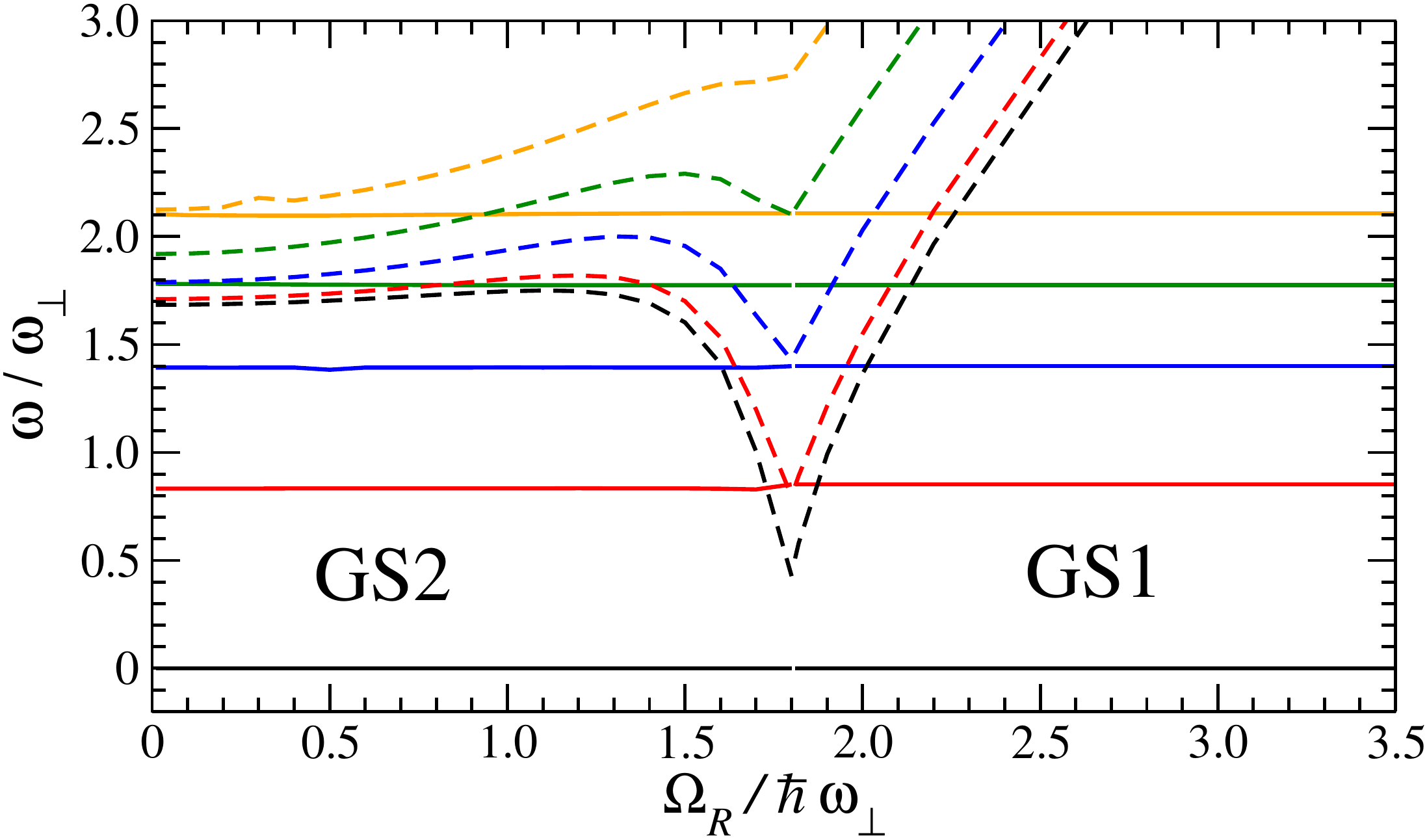}
	\caption{Lowest lying azimuthal ($n_\perp=0$) excitations of the ground state in the ring trap, for $g_{12}=1.5g$. Solid lines show the density mode, while dashed lines represent the spin-density mode. From bottom to top the solid and dashed lines correspond to the modes with $\ell=0,-1,-2,-3,-4$.}
	\label{Fig2}
\end{figure}

The lowest lying excitation frequencies of the ground state across the phase transition are shown in Fig.~\ref{Fig2}. For these modes there are small hybridization effects, and they can still be recognized as density and spin modes. 
In the limit of small $\Omega_R$ they tend to majority and minority component modes, respectively. The density modes (solid lines) keep almost constant values as $\Omega_R$ increases for the low lying modes shown in the figure, but they seem to show avoided crossings at higher quantum numbers (not shown). Instead, the spin modes vary strongly with $\Omega_R$. In particular, the $\ell=0$ spin mode (gap in the homogeneous case) drives the phase transition, which takes place around $\Omega_c\sim1.8\,\hbar\omega_\perp$, in agreement with the global polarization plot, Fig.~\ref{Fig1}. Using Eq.~\eqref{EqOmegac}, the critical value for the maximum density is predicted at $\Omega_c\sim 2.1\,\hbar\omega_\perp$. 

\section{Stability of persistent currents in GS1}\label{SecGS1}

In the GS1 phase, the criterion of stability of persistent currents is very similar to that of single component condensates, and it is related to energetic instabilities of the flow in the azimuthal direction when its velocity exceeds the Landau critical velocity.  
In the homogeneous ($V=0$) system at rest, the density and spin modes are given, respectively, by  \cite{Abad2013}
\begin{align}
	 (\hbar\omega_d)^2=& \frac{\hbar^2k^2}{2 m}\left(  \frac{\hbar^2k^2}{2 m} + (g+g_{12})n \right)\label{EqWDen}\\
	 (\hbar\omega_s)^2= &\frac{\hbar^2k^2}{2 m}\left(  \frac{\hbar^2k^2}{2m} + (g-g_{12})n +4\Omega_R\right) +\nonumber\\
	&+ 2\Omega_R\left[ (g-g_{12})n+2\Omega_R \right] \ .\label{EqWSpin}
\end{align}
Notice that the spin mode supports a dynamical instability, which occurs at the transition point between the GS1 and GS2 phases, see Eq.\eqref{EqOmegac}.
In the presence of currents, the excitation frequencies acquire a shift due to the Doppler effect $\omega'\to \omega+k v$, with $k$ the wave vector and $v$ the superfluid velocity. When the Doppler-shifted frequency fulfills $\omega'=0$ for some value of $k$, the currents become energetically unstable. 
This happens when the flow velocity reaches either the speed of sound of the density mode or the Landau velocity corresponding to the spin mode, respectively given by
\begin{align}
	&c_d = \sqrt{\frac{(g+g_{12})n}{2 m}}\,,\label{EqSound}\\
	&v_L^{(s)} \equiv\min[\omega_s(k)/k]= \sqrt{\frac{\hbar\omega_J}{m} + \frac{(g-g_{12})n+4\Omega_R}{2m}}\ ,\label{EqLandau}
\end{align}
where we have introduced the gap frequency (related to the internal Josephson frequency in the small amplitude limit),
\be
\hbar\omega_J = \sqrt{2\Omega_R\left[ (g-g_{12})n+2\Omega_R \right]} \ .\label{EqGap}
\ee
Depending on the values of the parameters, instabilities will be driven by density or spin-density excitations. In the absence of any polarization these two excitation channels are completely decoupled. 

If there is rotational symmetry, the excitation spectrum in the azimuthal direction of the ring trap is very similar to the Bogoliubov spectrum in free space, the main difference being that now the quasimomenta and the flow velocities are quantized due to the periodicity inherent to the ring trap. These two quantities can be written, respectively, in terms of the quantum number $\ell$ characterizing the azimuthal excitations, $k=2\pi \ell/R$, and the winding number $\kappa$ characterizing the persistent current, $v=\hbar\kappa/mR$, with $R$ the radius of the ring. In addition, the transverse degrees of freedom reduce the excitation frequency by a factor up to $\sqrt{2/3}$ in the Thomas-Fermi limit \cite{Abad2014}.

In the following we will analyze the decay of the currents in each of the modes in more detail, using numerical results obtained from BdG and GP.

\subsection{Instability driven by the density mode}

\begin{figure}\centering
	\includegraphics[clip=true,width=\linewidth]{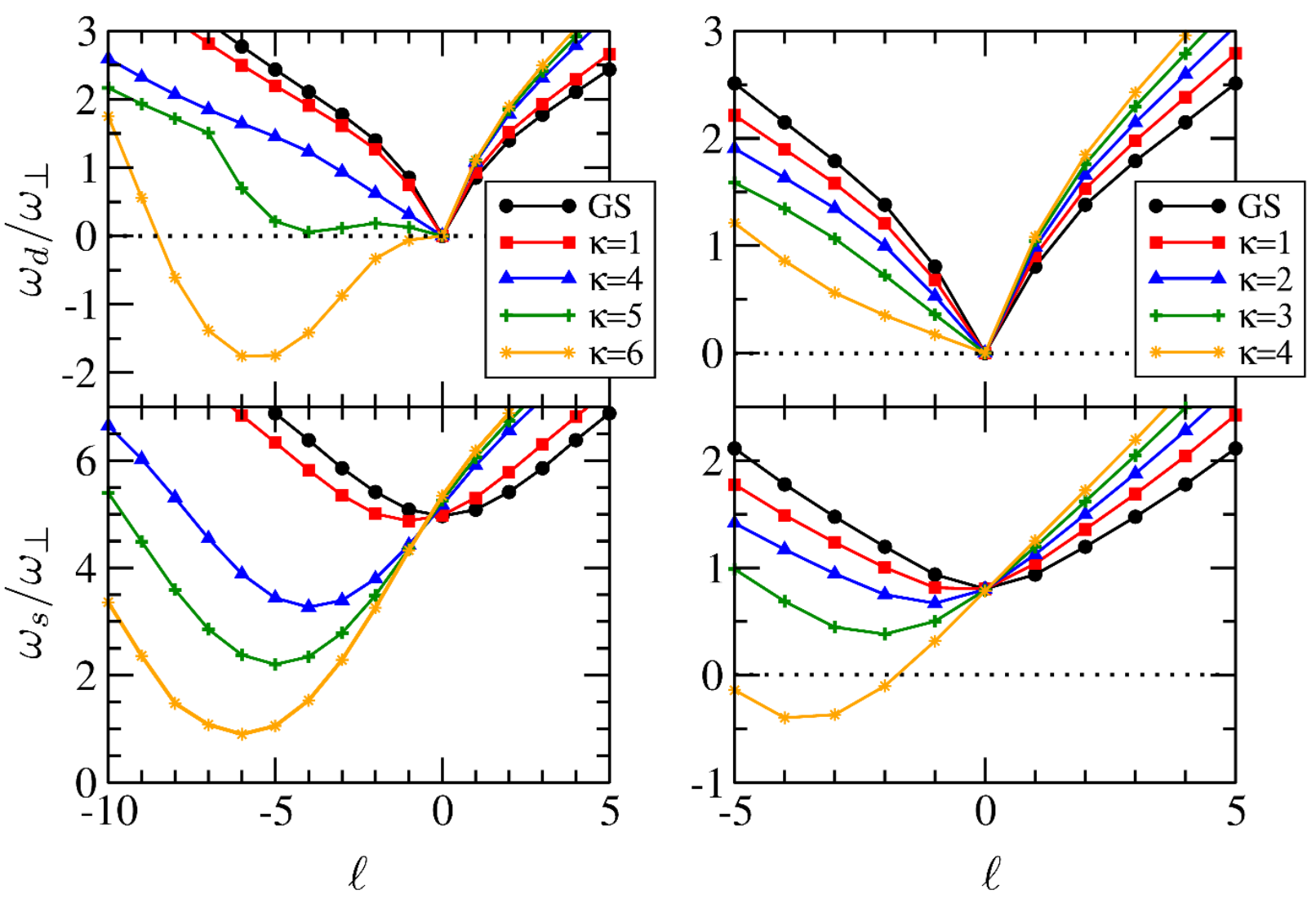}
	\caption{ Spectrum corresponding to azimuthal modes in GS1. Left panels: instability driven by the density mode, for $g_{12}=1.5g$, $\Omega_R=3.5\hbar\omega_\perp$; Right panels: instability driven by the spin mode, for $g_{12}=0.5g$, $\Omega_R=0.1\hbar\omega_\perp$. Top panels show the spectrum of density excitations, $\omega_d$, while the bottom panels show the spectrum of spin excitations, $\omega_s$. Lines have been drawn as a guide to the eye.}\label{Fig3}
\end{figure}

For large values of $\Omega_R$ the gap in the spin mode is so large that it is the density mode that becomes energetically unstable first. 
The left panel of Fig.~\ref{Fig3} shows examples of excitation spectra in this regime, for different values of $\kappa$. 
The upper and lower panels correspond, respectively, to the density and spin-density modes and the excitation frequencies are plotted as a function of $\ell$.  The case $\kappa=0$ corresponds to the ground state. For $\kappa>0$ we can see that the dispersion relations bend due to the Doppler shift and $\omega_d$ becomes negative for high enough $\kappa$. 
From Eq.~\eqref{EqSound}, we find that the sound velocity written in dimesionless form takes the value $\sqrt{2/3}c_d Rm/\hbar\sim4.5$, which is in agreement with the numerical results. The radius $R$ has been taken as that of the maximum density. Notice that, contrary to the homogeneous case, the gap in the spin mode shows a dependence on the winding number. This is a result of the density dependence on the winding number in a ring trap (experimentally studied in \cite{Moulder2012} for a single component condensate).

For high enough winding number, close to instability, a roton-like structure is seen to appear in the spectrum of the density mode. Its appearance has the effect of shifting the critical velocity to a finite wave-vector, and thus to lower its critical value compared to the homogeneous result Eq.~\eqref{EqSound}. 
The roton-like structure is not inherent of two-component systems but has been seen to appear as well in single components~\cite{Dubessy2012} and spinor $F=1$ systems~\cite{Makela2013}. It has been argued in \cite{Dubessy2012} that it comes from a coupling of the Bogoliubov mode to a surface mode \cite{Pethick,Anglin2001} localized in the internal surface of the ring. We have found that the roton minimum is more pronounced when the system is in the Thomas-Fermi limit in the radial direction, while it disappears when the system tends to a gaussian wave function. Indeed, in the limit of vanishing interactions the GP equations become Schr\"odinger equations and, due to the absence of the nonlinearity, the finite $\kappa$ states cannot break their symmetry and decay by shedding vortices.

\begin{figure}
	\includegraphics[clip=true,width=\linewidth]{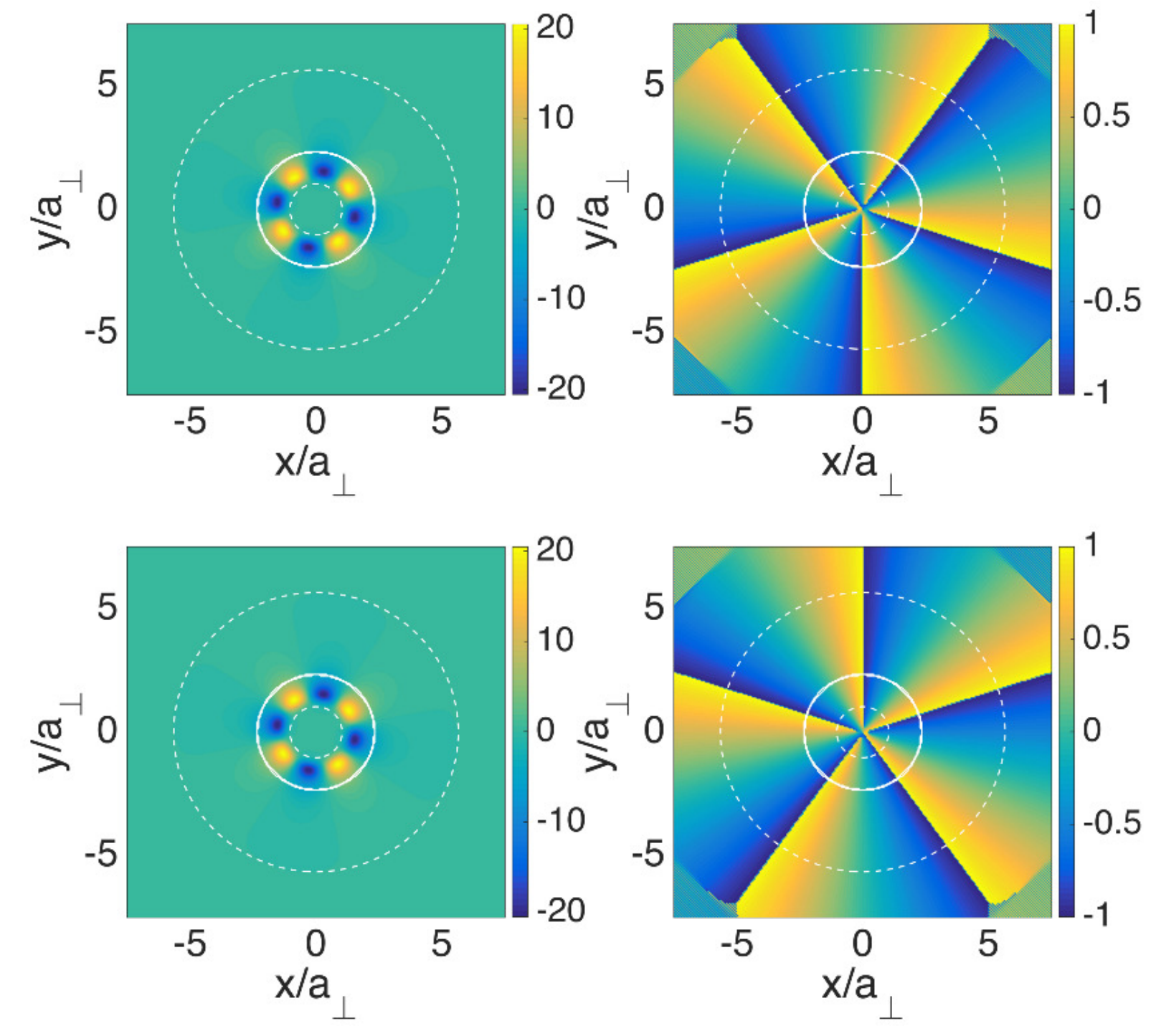}
	\caption{Wave functions $\Psi_1$ and $\Psi_2$ at time $t=0.1\,\omega_\perp^{-1}$ and for $\varepsilon=0.01$, Eq.~\eqref{EqPerturb}, perturbed with the mode with symmetry $n_\perp=0$, $\ell=-4$, in the case with $g_{12}=1.5g$, $\Omega_R=3.5\hbar\omega_\perp$ and $\kappa=5$. The left panels show $|\Psi_\alpha|^2-|\psi_\alpha|^2$ at first order in $\delta\Psi_\alpha$ (units of $a_\perp^{-2}$), while the right panels show the phase of $\Psi_\alpha$ (units of $\pi$). As a guide to the eye, the contours $\rho_\text{max}$ (solid) and $\rho_\text{max}/100$ (dashed) have been drawn.}\label{Fig4}
\end{figure}

To better understand the nature of the roton mode in the two-component system, we have looked at how the modes affect the wave functions $\Psi_\alpha(\mb{r},t)$, following Eq.~\eqref{EqPerturb}.
An example is shown in Fig.~\ref{Fig4} for an excitation corresponding to the roton-like minimum ($n_\perp=0$, $\ell=-4$) on top of the wavefuction with $\kappa=5$. The left panels show the density difference $|\Psi_\alpha|^2-|\psi_\alpha|^2$ (at first order in $\delta\Psi_\alpha$), while the right panels show the phase of $\Psi_\alpha$.
From the figure one can clearly see that this perturbation belongs to the density mode (densities in $1$ and $2$ components are in phase). 
Notice the $\pi$-phase difference between the $\delta\Psi_1$ and $\delta\Psi_2$; this is just the $\pi$-phase difference of the reference state, $\psi_\alpha$. This mode is clearly localized at the internal surface of the condensate.

We have found that, in our configuration, for $-7\le\ell\le-1$ the excitations of the $\kappa=5$ state are localized in the inner surface, while for $\ell>0$ and $\ell<-7$ they correspond to external surface excitations. Instead, for low values of $\kappa$ that do not show a roton minimum the excitations tend to be bulk excitations for low $\ell$ and external surface excitations at large $\ell$. As $\kappa$ is increased the low $\ell$ excitations become more localized at the internal surface. This seems to indicate that as the energetical instability is approached, the excitations tend to localize at the inner surface of the condensate at the same time that a roton mode with a minimum at finite $\ell$ is formed in the spectrum. When the system becomes unstable these modes lead to the nucleation of vortices, which later cause the phase-slips associated with the decay of the persistent current. Notice that the nucleation of vortices comes from a symmetry breaking of the pure finite $\kappa$ state, which is only possible due to the nonlinearity of the GP equations. On the other hand, the nucleation of vortices can only happen with finite wavelength excitations (that is, not by sound-like excitations), and the only way to achieve these is by a roton structure.

\begin{figure*}\centering
	\includegraphics[clip=true,width=0.41\linewidth]{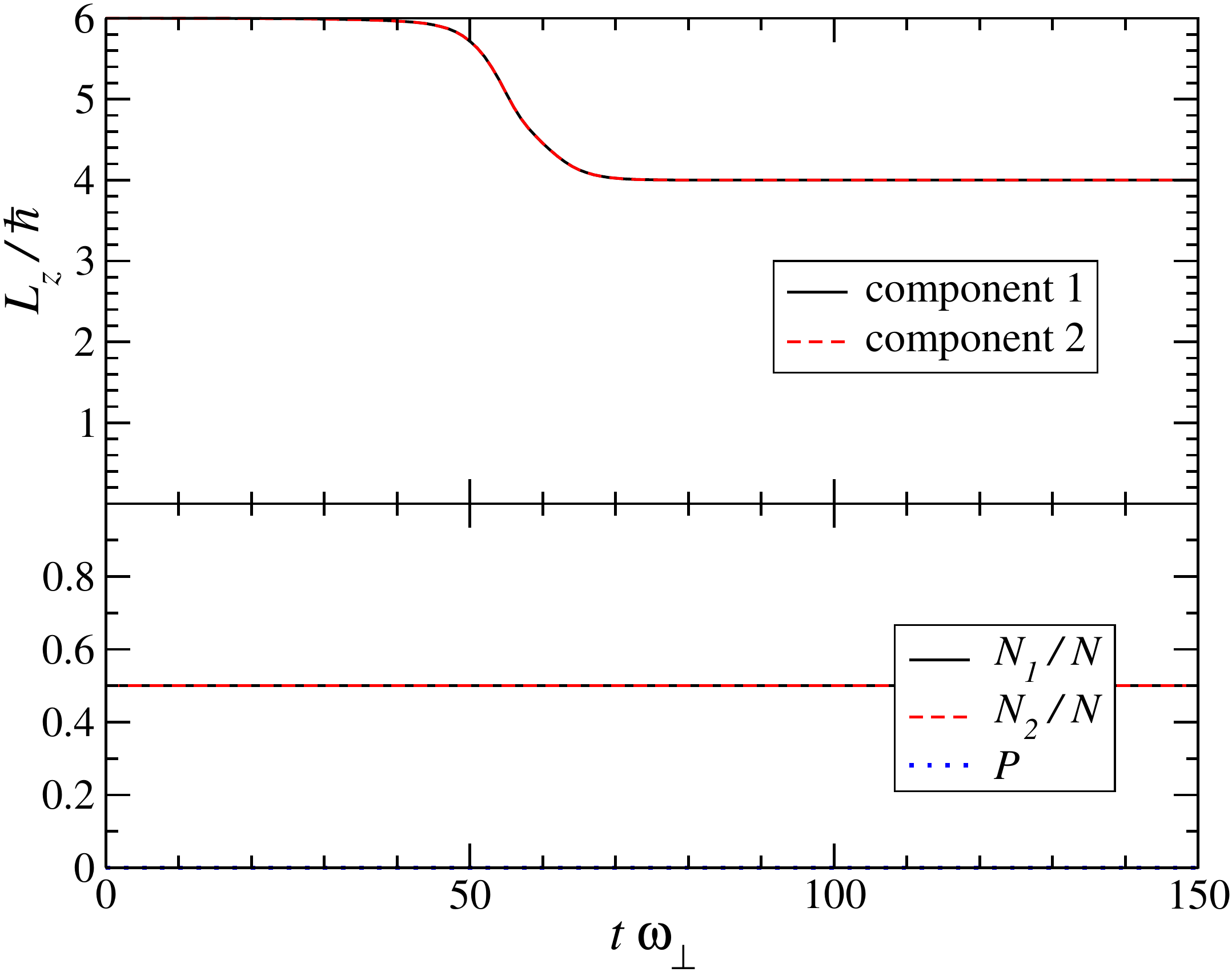}%
	\includegraphics[clip=true,width=0.19\linewidth]{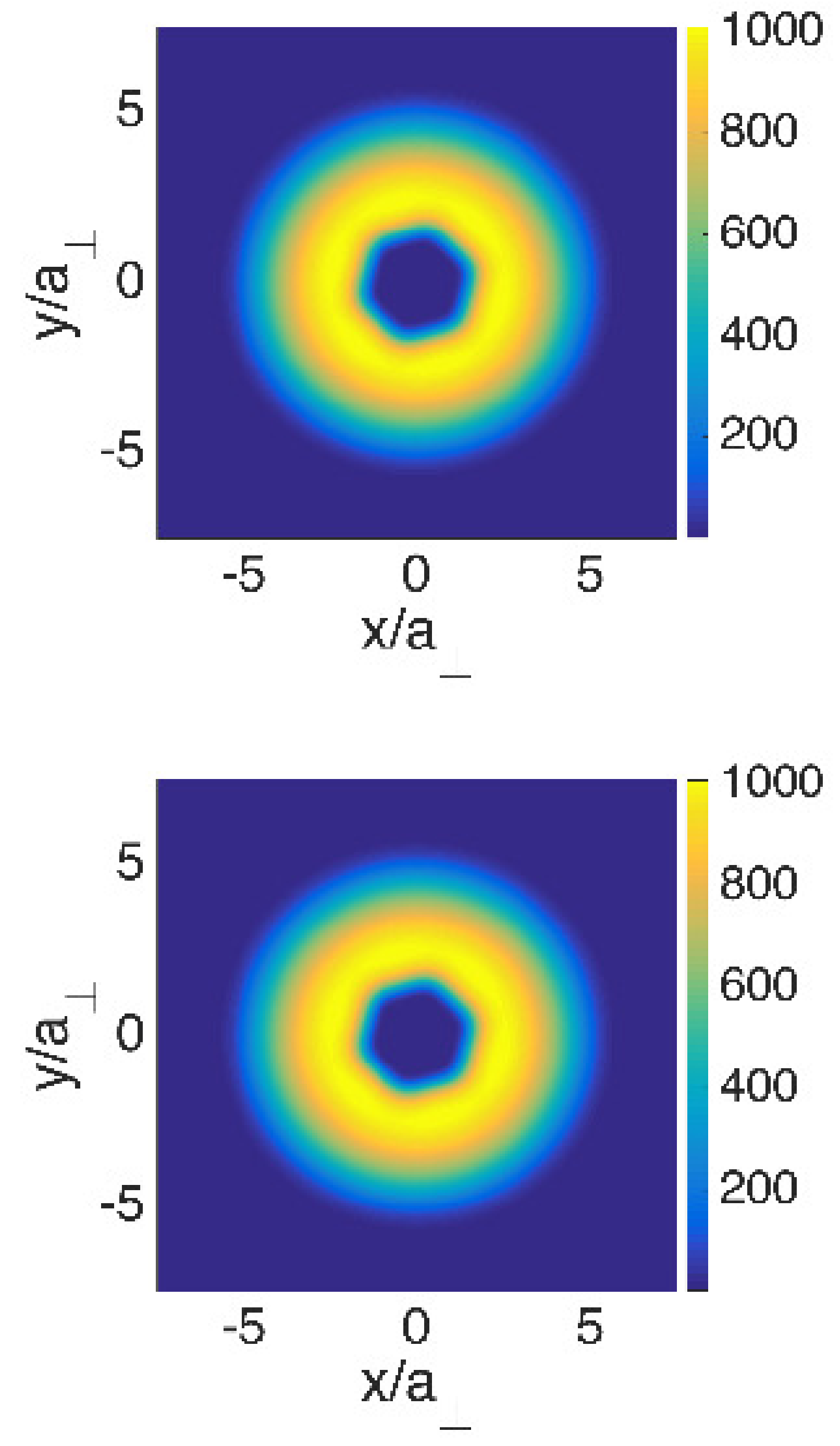}%
	\includegraphics[clip=true,width=0.19\linewidth]{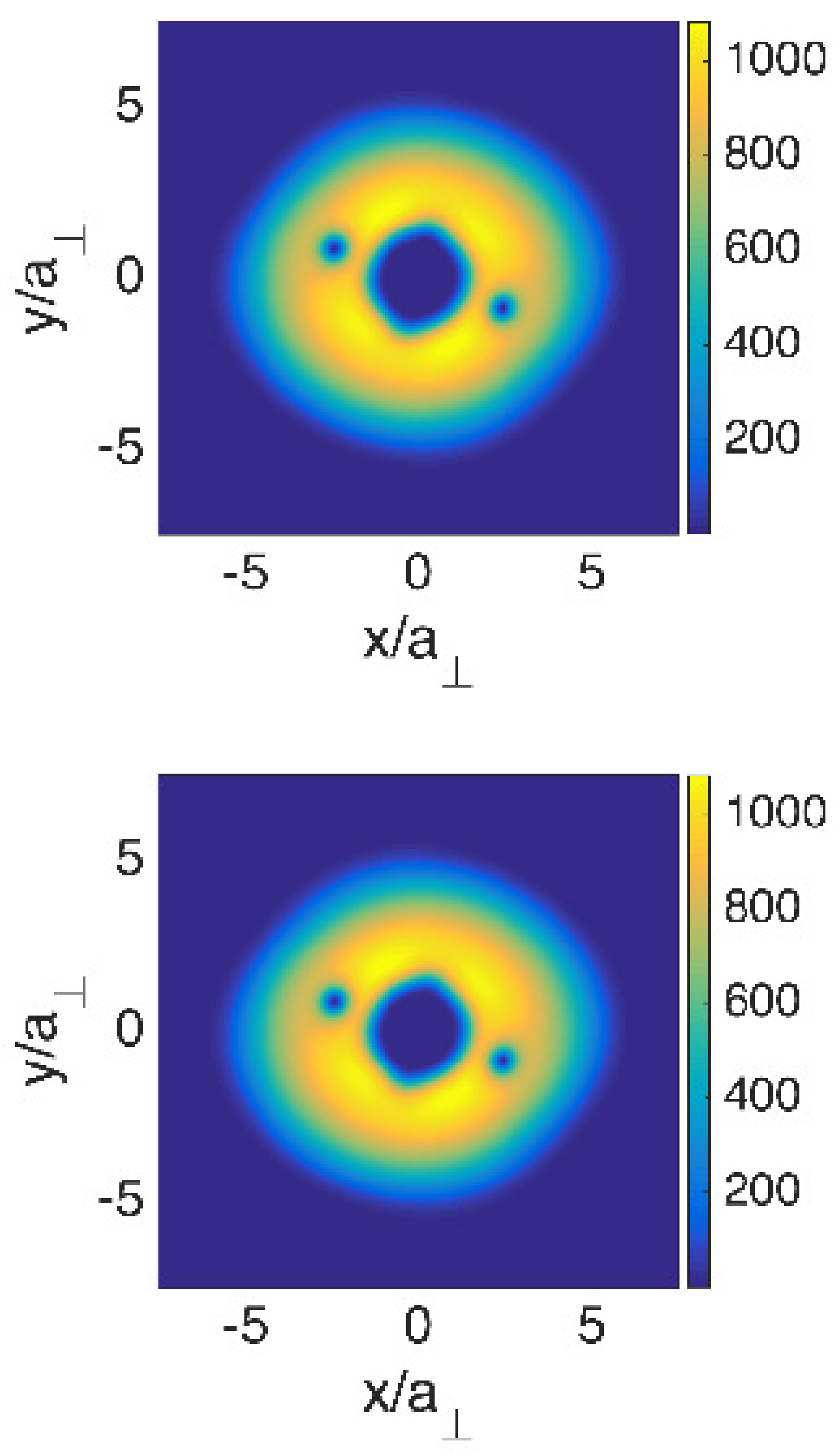}%
	\includegraphics[clip=true,width=0.19\linewidth]{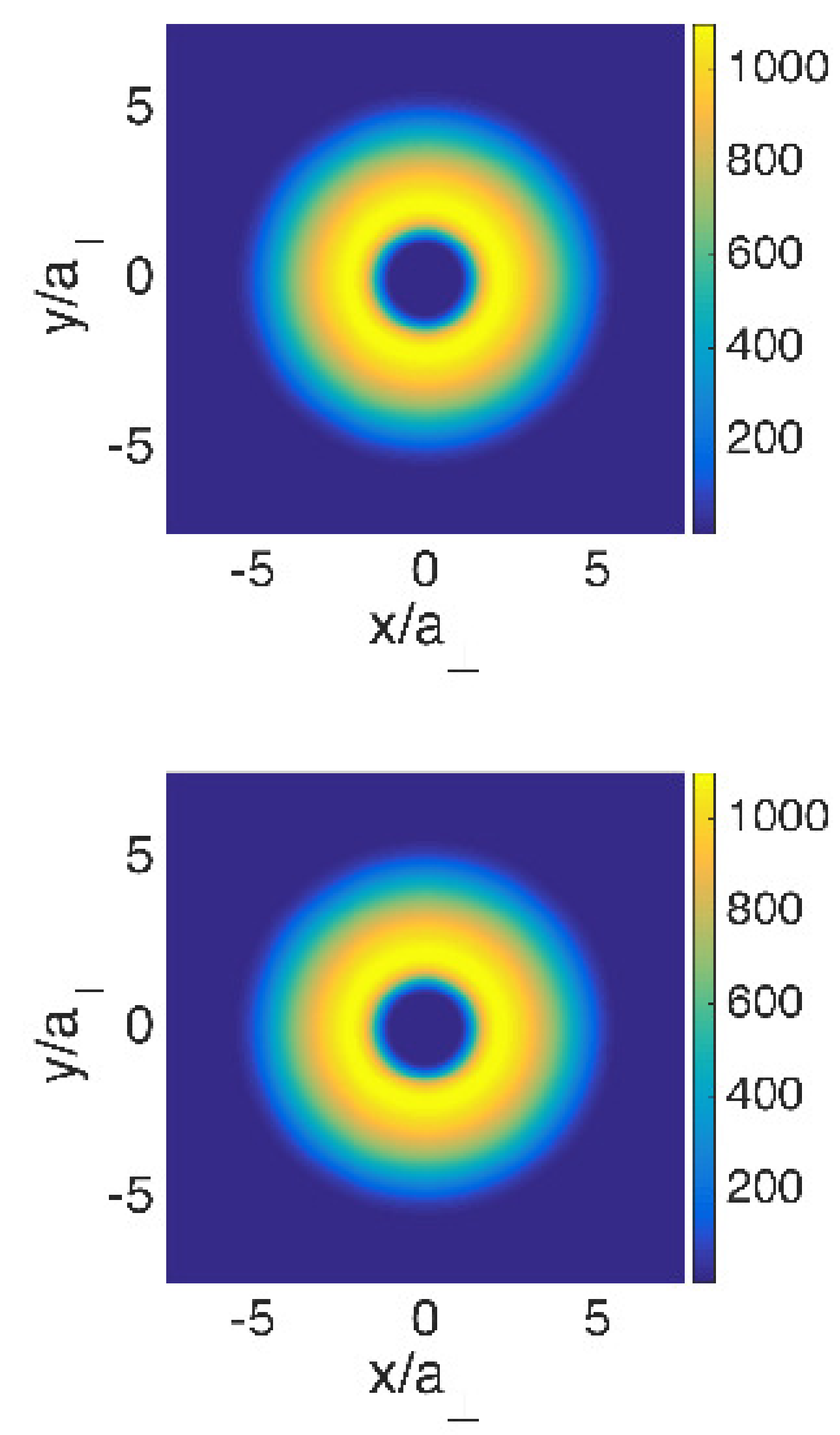}
	\caption{Dissipative dynamics of the decay of persistent currents with $\kappa=6$ in a GS1 configuration with $g_{12}=1.5g$ and $\Omega_R=3.5\,\hbar\omega_\perp$. The initial state is built using Eq.~\eqref{EqPerturb} with $\varepsilon=0.01$, using the eigenmode with $\ell=-6$ with density character. Left panels show the angular momentum and the numbers of particles, respectively. Density plots (in harmonic oscillator units) show components 1 (top row) and 2 (bottom row) at times: $t=20\,\omega_\perp^{-1}$ (left), $t=52\,\omega_\perp^{-1}$ (middle), and $t=94\,\omega_\perp^{-1}$ (right).}\label{Fig5}
\end{figure*}

To explore the decay of the superflow in the long-time and non-linear regimes we have evolved Eqs.~\eqref{Eq:GP1}--\eqref{Eq:GP2} in real time using Eq.~\eqref{EqPerturb} with $\varepsilon=0.01$ as the $t=0$ wave function, adding the small dissipative term $\gamma$ (see Sec.~\ref{SecTheoryGP}). 
The time evolution of the angular momentum per particle, $L_z$, and the polarization $P$ are shown in Fig.~\ref{Fig5} together with snapshots of the density taken at different times, for the case with $g_{12}=1.5\,g$ and $\Omega_R=3.5\,\hbar\omega_\perp$. It is clear from the figure that both components show an identical evolution, and the decay happens as for a single component. The initial $\ell=-6$ excitation evolves into the nucleation of two vortices in the internal surface, which spiral out of the condensate producing thus the phase slips that drive the decay of the flow. The final state has $\kappa=4$, which is stable. Notice that the two vortices happen at exactly the same place in both components.

\subsection{Instability driven by the spin-density mode}

Close enough to the GS1-GS2 phase transition, that is for small enough values of $\Omega_R$ but still $\Omega_R>\Omega_c$, the gap in the spin mode is small, and the decay of the persistent currents is driven by out-of-phase excitations in the azimuthal direction. 
An example of dispersion relations in the case where the lowest mode is the spin-density mode is shown in the right panels of Fig.~\ref{Fig3}. When the flow velocity exceeds $v_L^{(s)}$, see Eq.~\eqref{EqLandau}, the spectrum becomes negative at finite $\ell$ and the persistent currents are no longer stable. In the case of the figure, the spin Landay velocity written in dimesionless form takes the value $\sqrt{2/3}v_L^{(s)}Rm/\hbar\sim 3.2$, which agrees well with both imaginary time simulations and BdG.

\begin{figure}
	\includegraphics[clip=true,width=\linewidth]{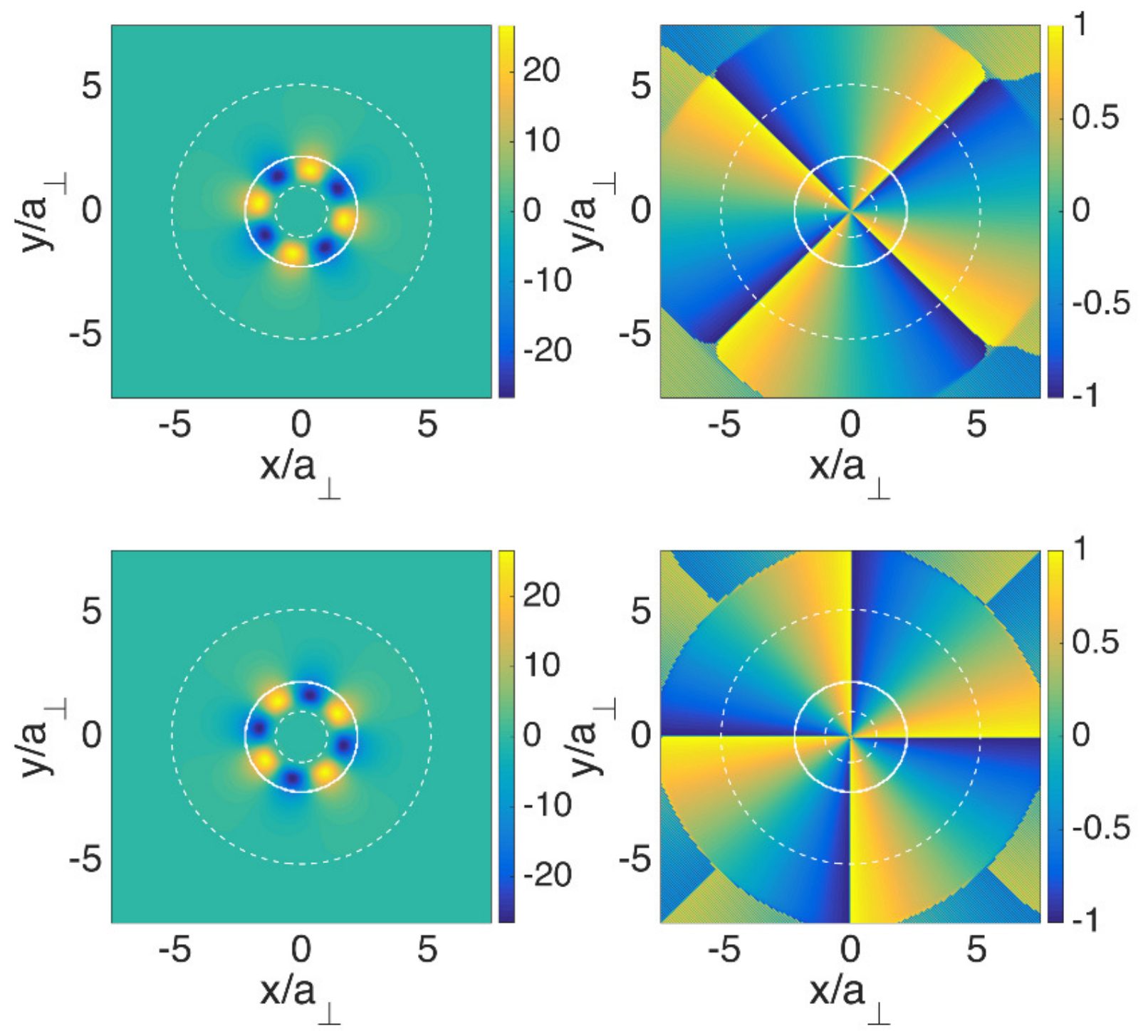}
	\caption{Wave functions $\Psi_1$ (top) and $\Psi_2$ (bottom) at time $t=0.1\,\omega_\perp^{-1}$ and for $\epsilon=0.01$, Eq.~\eqref{EqPerturb}, perturbed with the mode with symmetry $n_\perp=0$, $\ell=-4$, in the case with $g_{12}=0.5g$, $\Omega_R=0.1\,\hbar\omega_\perp$ and $\kappa=4$. The left panels show $|\Psi_\alpha|^2-|\psi_\alpha|^2$ at first order in $\delta\Psi_\alpha$ (units of $a_{\perp}^{-2}$), while the right panels show the phase of $\Psi_\alpha$ (units of $\pi$). As a guide to the eye, the contours $\rho_\text{max}$ (solid) and $\rho_\text{max}/100$ (dashed) have been drawn.}\label{Fig6}
\end{figure}

It is interesting to note that the spin mode does not show any roton-like structure. This is due to the fact that, since the mode is gapped, the minimum of the excitation spectrum is already at $\ell<0$, allowing thus the appearance of excitations at finite values of $\ell$, necessary for the nucleation of vortices. 
When the dispersion relation approaches the energetic instability, the eigenvectors of the lowest lying excitations in the spin-density mode localize mostly at the internal surface, as happened in the density mode. 
As an example, Fig.~\ref{Fig6} shows the perturbed wave functions $\Psi_\alpha$ for the unstable mode with $n_\perp=0$ and $\ell=-4$ for $\kappa=4$ (see Fig.~\ref{Fig3}). One clearly sees that it corresponds to a spin-density mode localized at the inner surface of the condensate, where the perturbations in the two components appear out of phase. 

\begin{figure*}\centering
	\includegraphics[clip=true,width=0.41\linewidth]{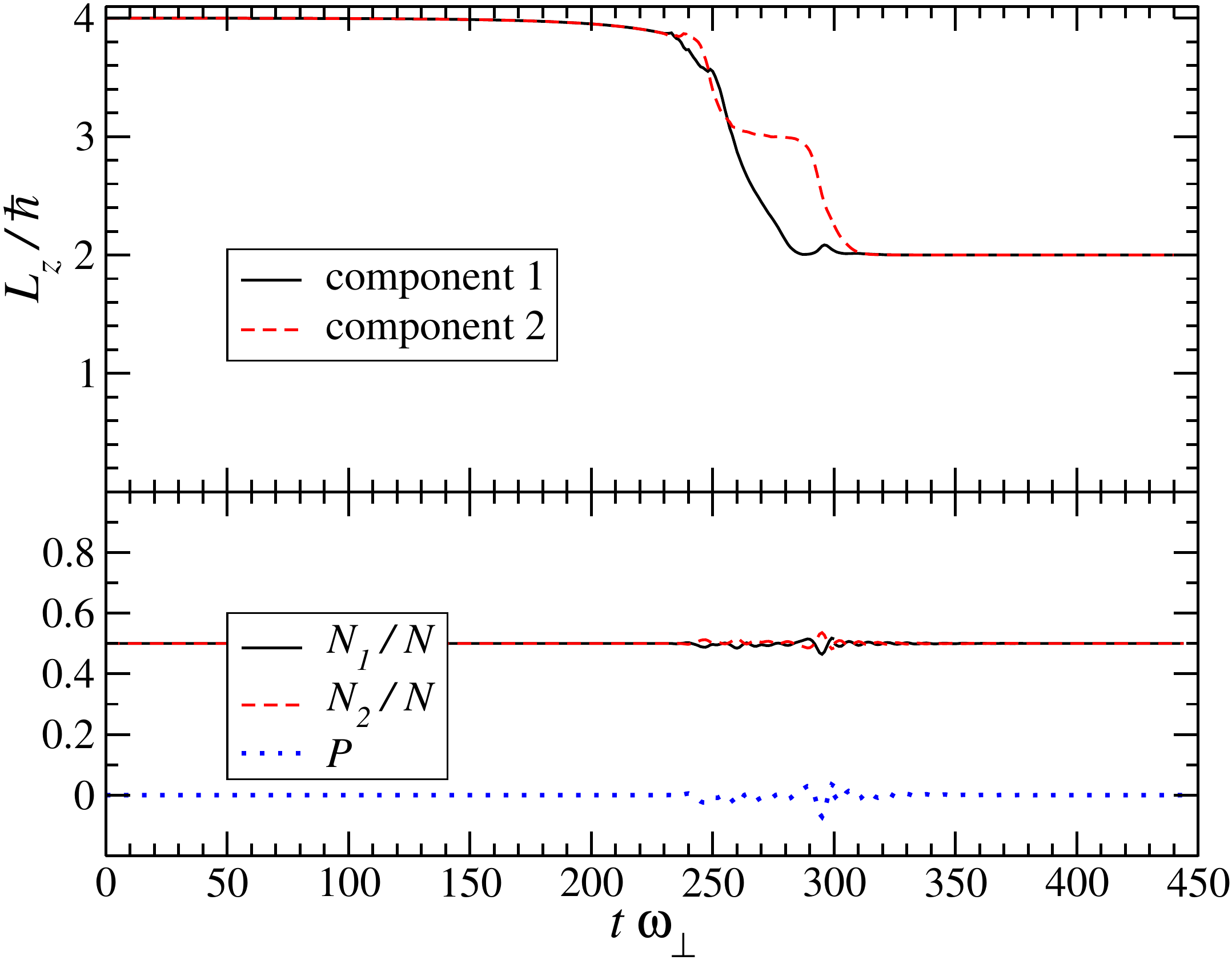}%
	\includegraphics[clip=true,width=0.19\linewidth]{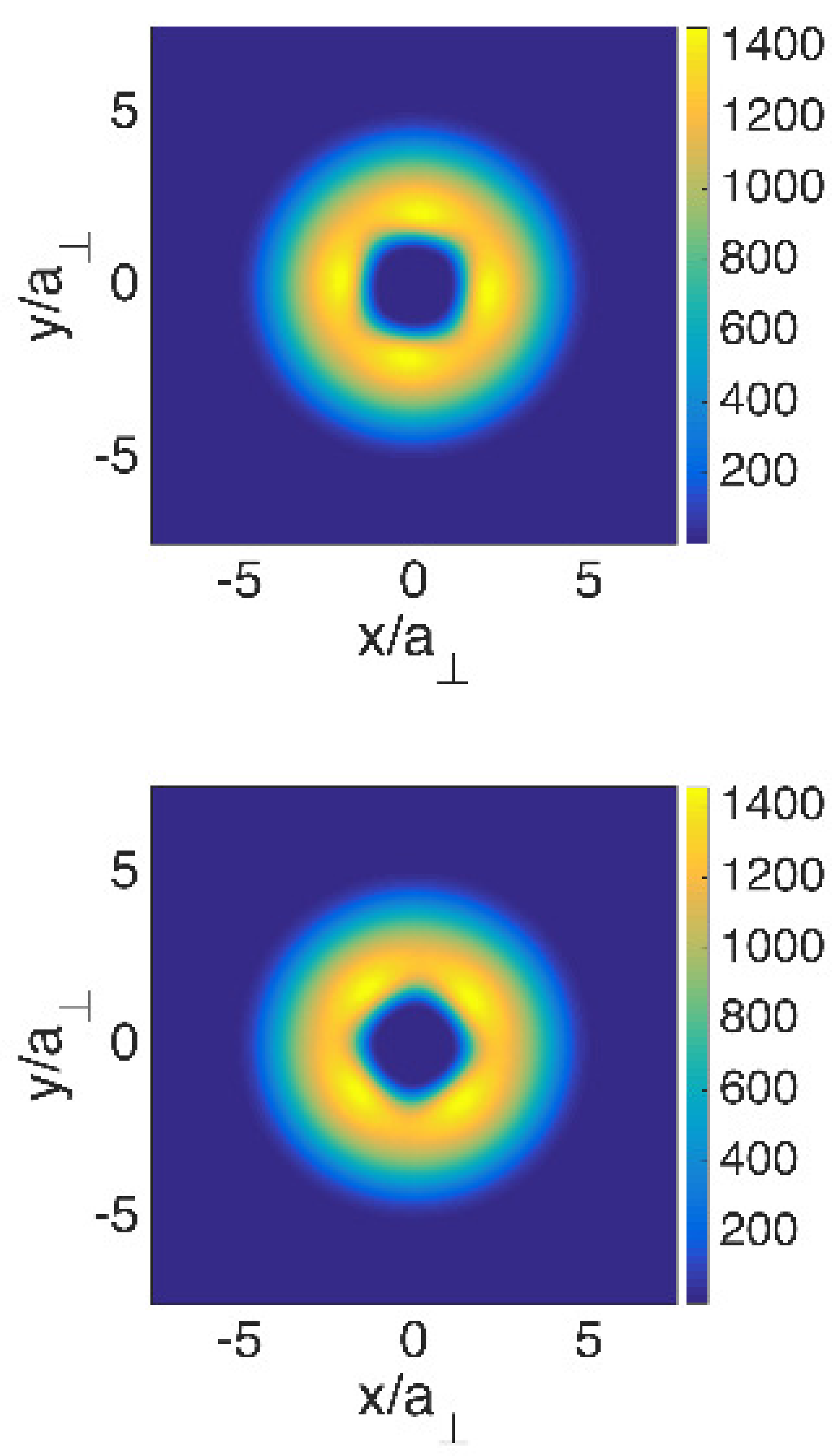}%
	\includegraphics[clip=true,width=0.19\linewidth]{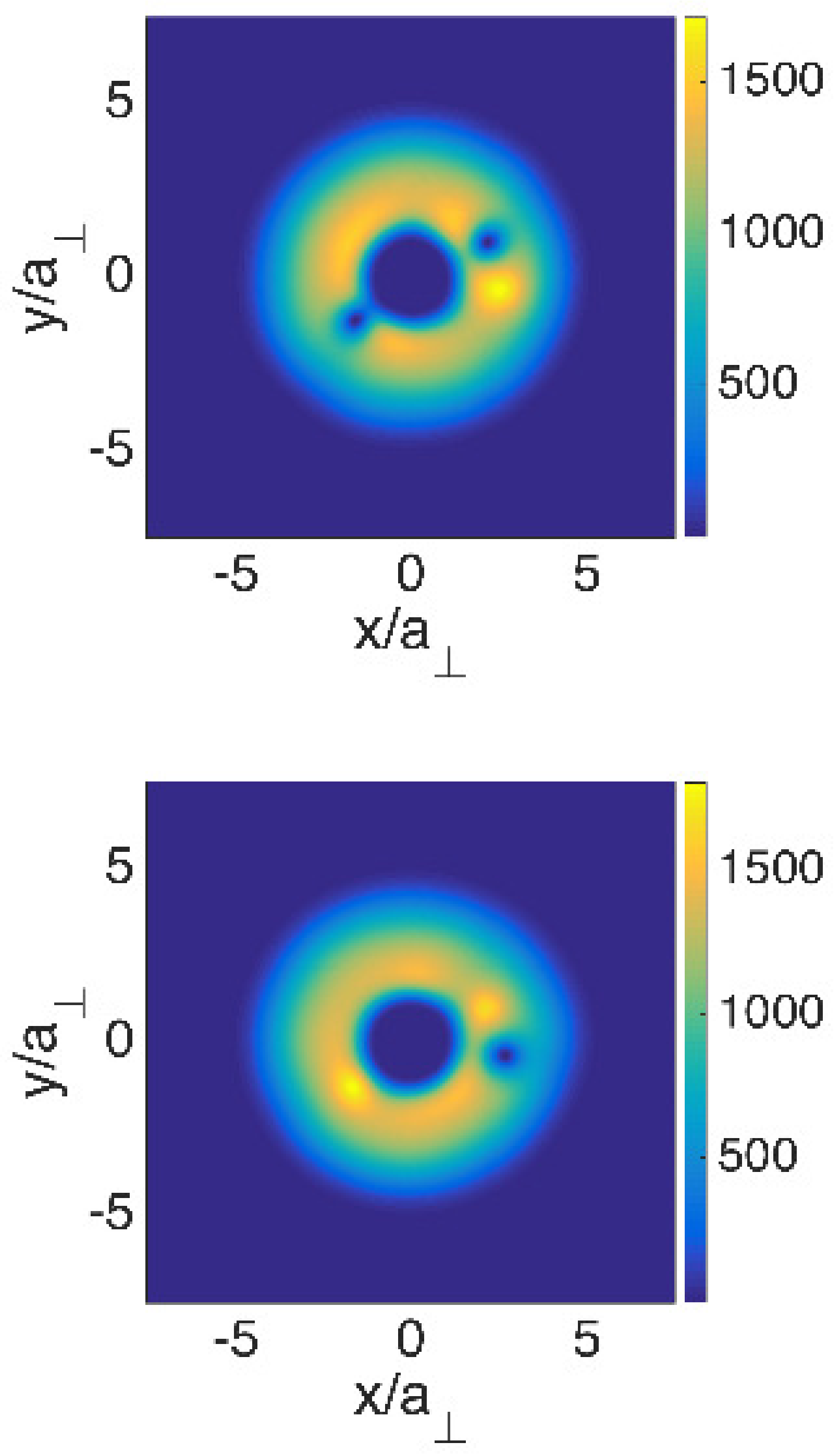}%
	\includegraphics[clip=true,width=0.19\linewidth]{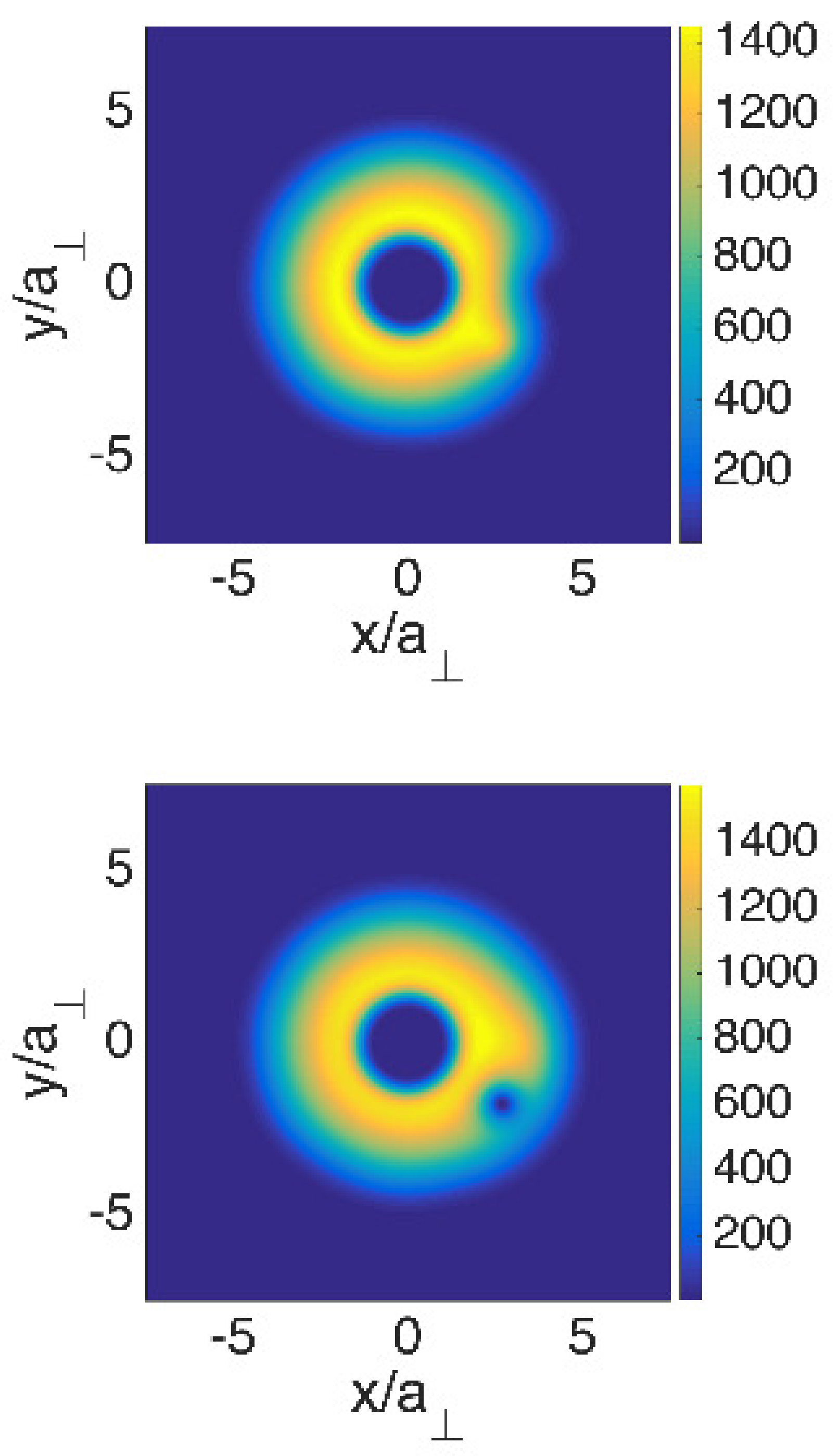}
	\caption{Dissipative dynamics of the decay of persistent currents with $\kappa=4$ in a GS1 configuration with $g_{12}=0.5g$ and $\Omega_R=0.1\,\hbar\omega_\perp$. The initial state is built using Eq.~\eqref{EqPerturb} with $\varepsilon=0.01$, using the eigenmode with $\ell=-5$ with spin character. Left panels show the angular momentum and the numbers of particles, respectively. Density plots (in harmonic oscillator units show components 1 (top row) and 2 (bottom row) at times: $t=142\,\omega_\perp^{-1}$ (left), $t=248\,\omega_\perp^{-1}$ (middle), and $t=290\,\omega_\perp^{-1}$ (right).}\label{Fig7}
\end{figure*}

To understand how the spin excitation drives the decay dynamics of the persistent currents, we have numerically evolved the GP Eqs.~\eqref{Eq:GP1} and \eqref{Eq:GP2} in real time with dissipation, with the initial wave function plotted in Fig.~\ref{Fig6}, for which $g_{12}=1.5g$, $\Omega_R=0.1\,\hbar\omega_\perp$, $\kappa=4$ and $\ell=-4$. The time evolution of the angular momentum and the polarization is shown in Fig.~\ref{Fig7} together with selected density snapshots during the evolution. In contrast to the decay driven by the density mode, we can see from the figure that the original out-of-phase excitation evolves into the nucleation of vortices between components that need not be correlated. As we can see from the middle snapshots, component 1 develops two vortices while component 2 only shows only one vortex. The two vortices (one in component 1 and the other in component 2), which in this plot appear on the right, move together during the dynamics, until they exit the condensate; they seem to form some kind of bound state. In contrast, the other vortex event exists separately in both components: component 1 expels the second vortex first (seen as well in the angular momentum evolution as a drop), and later component 2 follows, in a way that seems to bind its vortex with a ghost vortex of the first component, which is perceived in the figure as the irregular shape of component 1. This complicated dynamics is reflected as oscillations of the numbers of particles in each component, and in the polarization. Let us note that this simulation is an example of the kind of dynamics that could arise when the decay of the persistent currents is driven by spin-density excitations. It shows that phase-slip events can exist simultaneously in both components but also separately, in contrast to the decay driven by the density mode. The exact dynamical evolution and the final $\kappa$ state may depend on the initial conditions and the dissipation term $\gamma$.

\section{Stability of persistent currents in GS2}\label{SecGS2}

In the GS2, the conditions for stability of persistent currents are more complex. Firstly, the presence of currents lowers the value of the global polarization with respect to the ground state, that is the system with nonzero superflow is less polarized than the ground state. This implies that the energetic instability driving the decay of the currents cannot be understood as a simple Doppler shift, since the velocity shifts the system with polarization $P$ at rest to another configuration characterized by $P'<P$.  Secondly, the excitation modes (now hybridized due to the finite polarization) depend strongly on the value of $\Omega_R$ since the densities change with polarization. These two points reflect the fact that the GS2 is not a unique configuration, but each configuration with different $P$ is a different ground state. Instead, in the GS1 phase, there is a unique ground state characterized by equal density in both components and phase locking.

\begin{figure}\centering
	\includegraphics[clip=true,width=\linewidth]{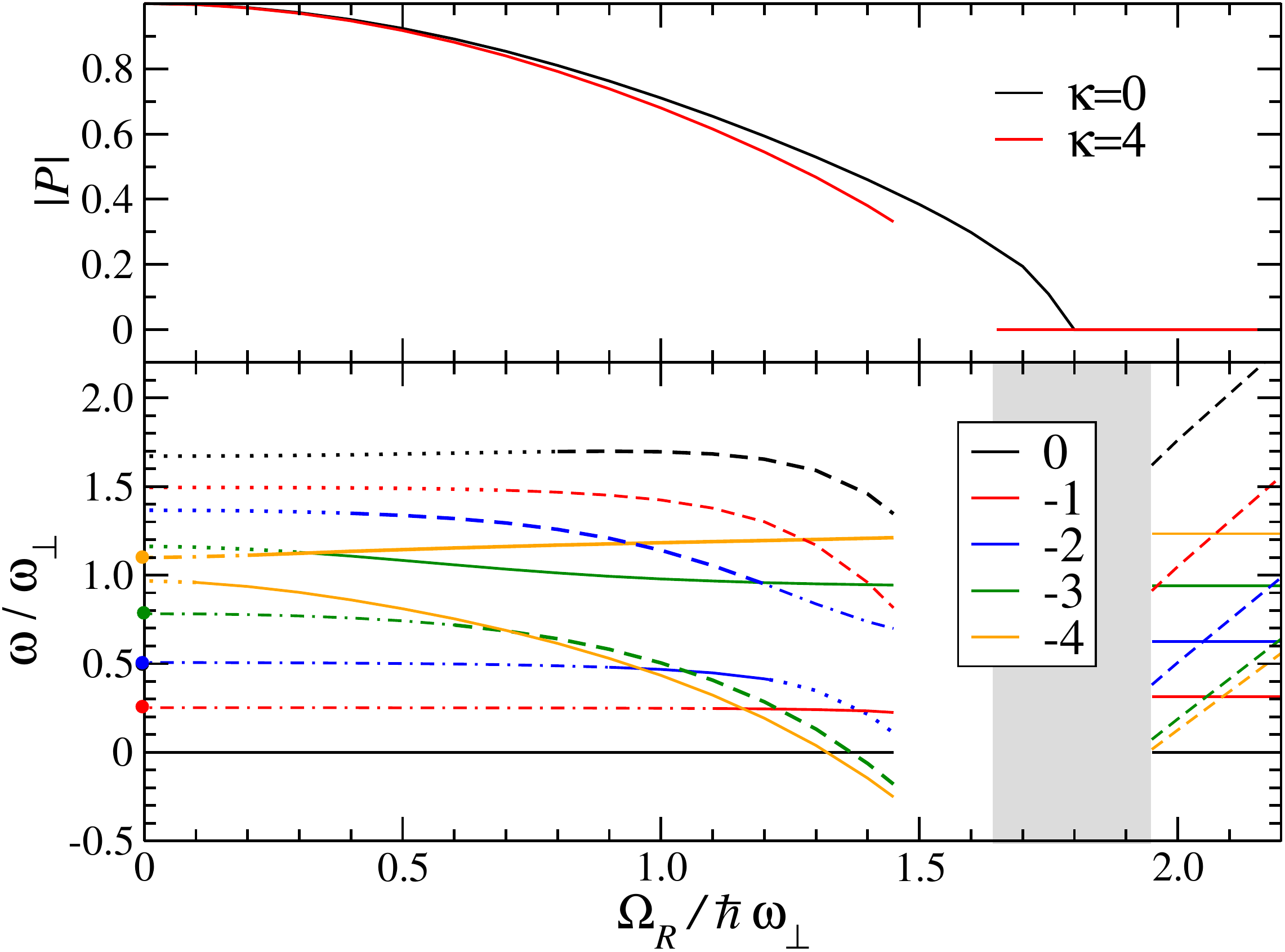}
	\caption{Top panel: Global polarization for $\kappa=0$ and $\kappa=4$ as a function of $\Omega_R$, for $g_{12}=1.5g$. Bottom panel: Excitation modes $\ell=0,...,-4$ of the $\kappa=4$ case as a function of the Rabi coupling. Solid lines correspond to modes with density character; dashed lines correspond to modes with spin character; dash-dotted lines correspond to modes with majority component character; dotted lines correspond to modes with minority component character. The shaded region corresponds to the states with broken rotational symmetry, and the blank region to states where the $\kappa=4$ current is highly unstable.}\label{Fig8}
\end{figure}

An example of the two points above is shown in Fig.~\ref{Fig8}, for a system with $g_{12}=1.5g$. The top panel compares the global polarization for $\kappa=4$ and $\kappa=0$ as a function of $\Omega_R$, while the bottom panel shows the  excitation energies of the modes with azimuthal quantum number $\ell=0,-1,-2,-3,-4$. For $\Omega_R\ge 1.95\,\hbar\omega_\perp$, the system is unpolarized (GS1) and the stability of persistent currents follows what has been discussed in Sec.~\ref{SecGS1}. For $\Omega_R<1.95\,\hbar\omega_\perp$, one of the spin modes becomes energetically unstable, but instead of decaying, the system undergoes an azimuthal symmetry breaking (see discussion below). This allows the system to remain unpolarized down to $\Omega_R\simeq1.65\,\hbar\omega_\perp$, but since the symmetry is broken the excitation spectrum cannot be easily classified in terms of $\pm\ell$ quantum numbers (shaded area in the bottom panel of Fig.~\ref{Fig8}). Below $\Omega_R\simeq1.65\,\hbar\omega_\perp$, there is a range of $\Omega_R$ (empty region in the figure) for which the current with $\kappa=4$ is highly unstable (numerically, this means that imaginary time cannot find a local minimum without any constraint). At lower $\Omega_R$ the system shows a regime of energetic instability (with preserved rotational symmetry) in the mode with mainly spin character. Finally, below $\Omega_R\simeq 1.3\,\hbar\omega_\perp$ the polarization increases with decreasing $\Omega_R$ and the persistent currents are stable. In the following we analyze these regimes in more detail.

Let us start with the persistent currents for $\Omega_R$ below the highly unstable region.
In this regime, the excitations can be described in terms of the $\pm\ell$ quantum numbers since the system shows rotational symmetry. However, the modes no longer correspond to pure density or spin-density modes as in the GS1, but they are now hybridized and the spectrum shows avoided crossings. 
In the limit $\Omega_R\ll\Omega_c$, where the system is highly polarized, the excitation modes are described in terms of excitations of the majority or the minority components. This character is shown in the figure as dash-dotted and dotted lines, respectively. The single-component character represents excitations for which the norm in one component is more than $10\,\%$ of the norm of the other component. Surprisingly, the single-component character is recovered for the $\ell=-2$ modes close to the unstable region. It is also interesting to note that for the $\ell=-3$ excitation the majority component character becomes spin character.

\begin{figure*}\centering
	\includegraphics[clip=true,width=0.41\linewidth]{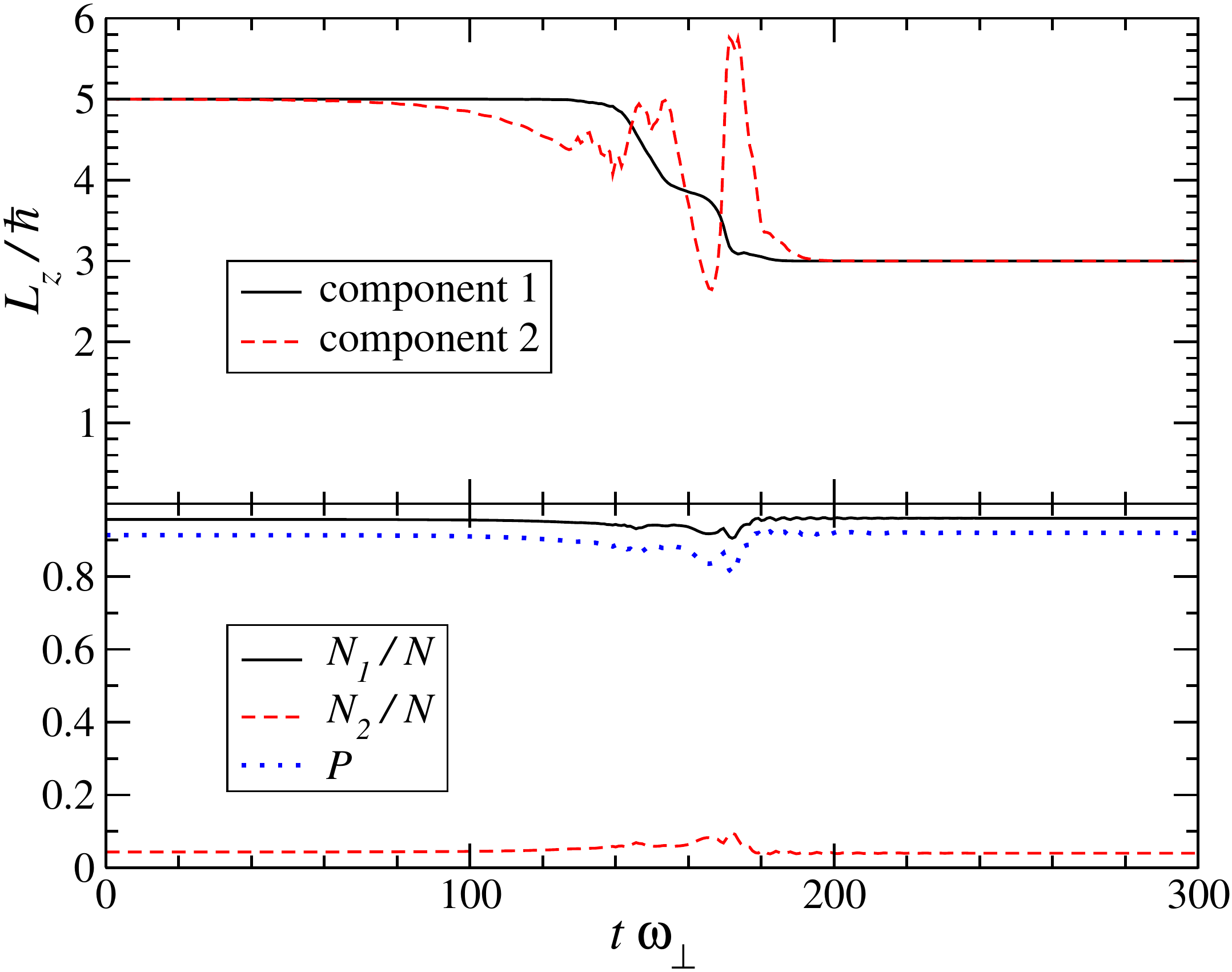}%
	\includegraphics[clip=true,width=0.19\linewidth]{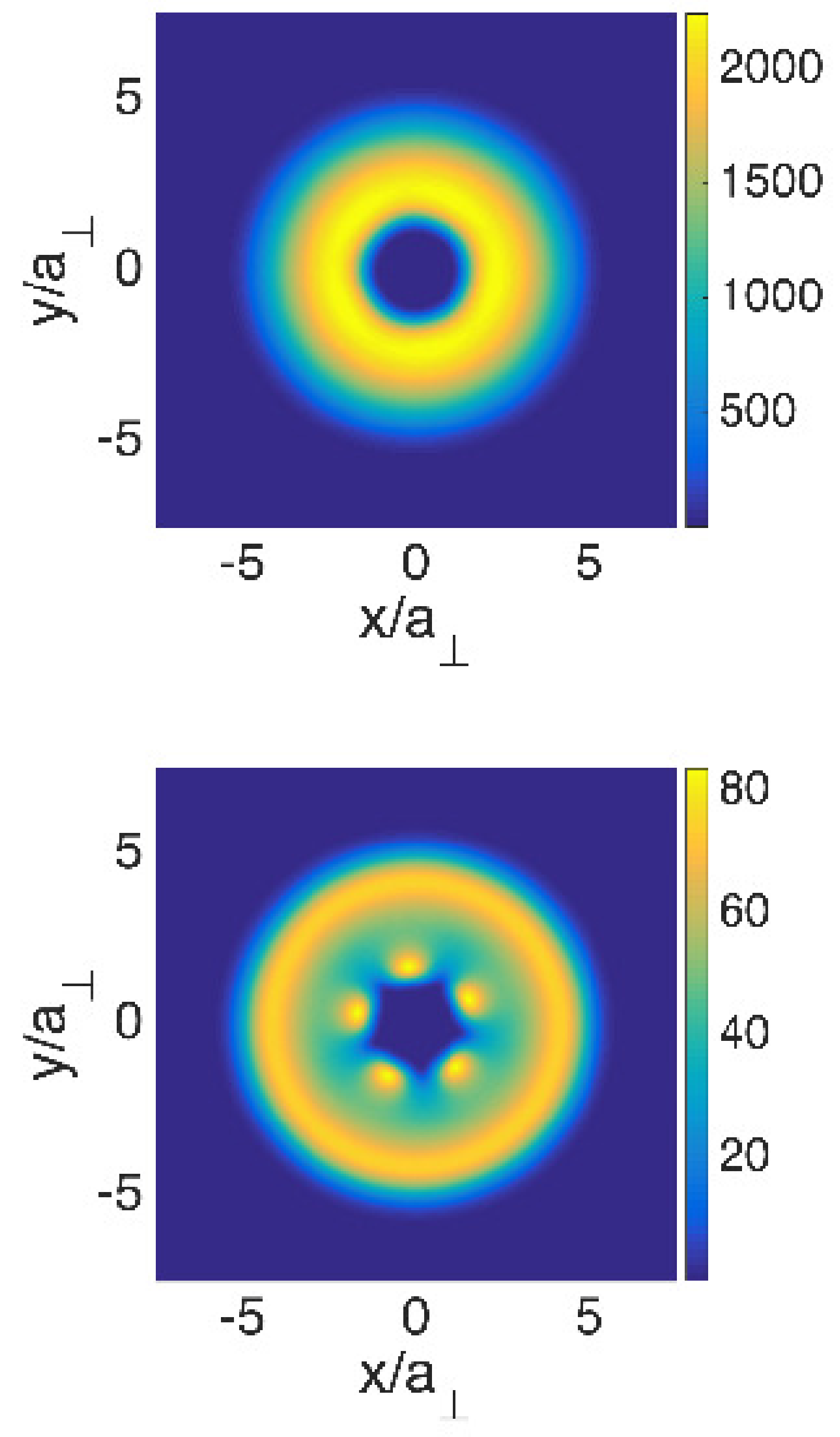}%
	\includegraphics[clip=true,width=0.187\linewidth]{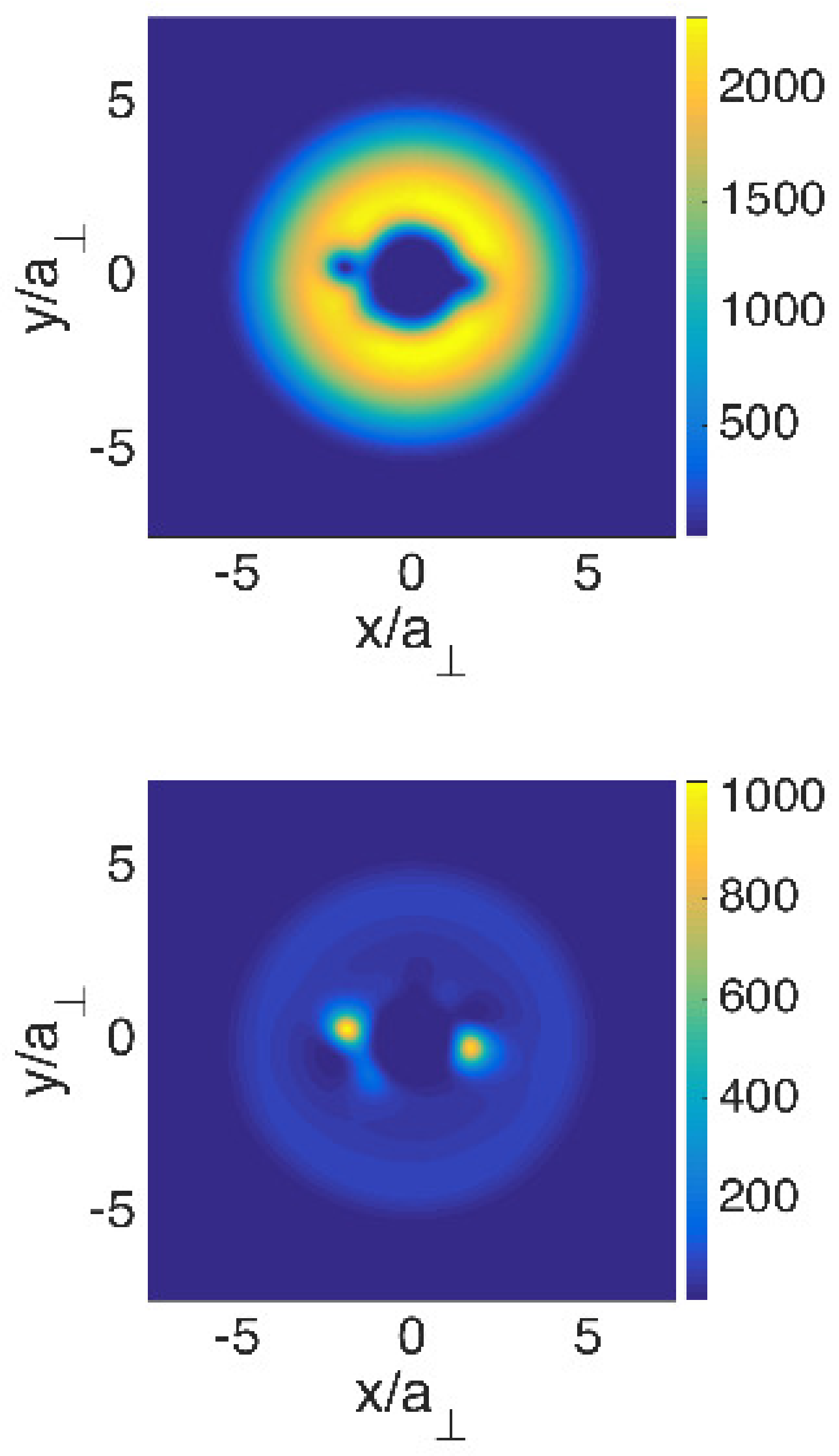}%
	\includegraphics[clip=true,width=0.187\linewidth]{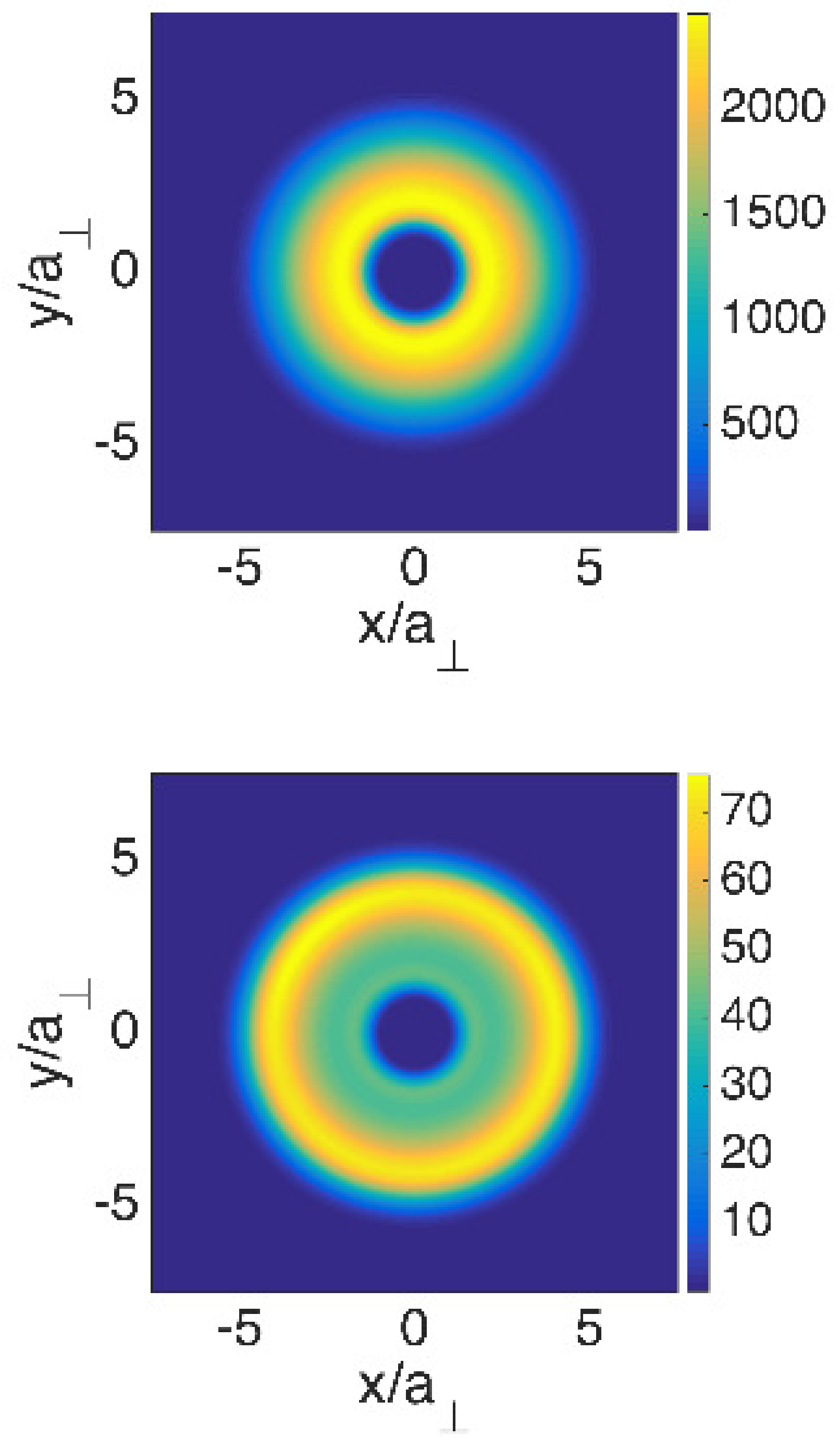}
	\caption{Dissipative dynamics of the decay of persistent currents with $\kappa=5$ in a GS2 configuration with $g_{12}=1.5g$ and $\Omega_R=0.5\,\hbar\omega_\perp$. The initial state is built using Eq.~\eqref{EqPerturb} with $\varepsilon=0.01$, using the eigenmode with $\ell=-5$ with minority component character (mainly spin-like). Left panels show the angular momentum and the numbers of particles, respectively. The density plots (in harmonic oscillator units) show the majority (top row) and minority (bottom row) components at times: $t=80\,\omega_\perp^{-1}$ (left), $t=140\,\omega_\perp^{-1}$ (middle), and $t=228\,\omega_\perp^{-1}$ (right). }\label{Fig9}
\end{figure*}

Figure~\ref{Fig9} shows the time evolution of the angular momentum per particle and the polarization together with snapshots of the decay dynamics of a persistent current in the GS2, corresponding to a state with winding number $\kappa=5$, $g_{12}=1.5g$, and $\Omega_R=0.5\,\hbar\omega_\perp$. In this case the flow in both the majority and the minority components (density-like and spin-like modes) is energetically unstable and decay is mostly driven by spin excitations. The initial excitation appears mainly on the minority component, but as time goes on it transfers to the majority component as well. As soon as the latter begins to allow vortices in the high density regions, the density in the minority component becomes trapped in the vortex cores without losing its coherence. This leads to the spikes of angular momentum of the minority component, as well as to exchange dynamics between the components. As happened when the decay in the GS1 was driven by spin excitations, the vortices in both components seem to form inter-component bound states, where the cores are very close to each other and move in pairs. When the vortices are finally expelled, the system regains equilibrium at a new value of the angular momentum.

As $\Omega_R$ increases and the polarization of the mixture decreases, the spin and density characters of the excitation modes are recovered. Avoided crossings may take place between modes with density and spin characters and the same quantum number $\ell$. These avoided crossings occur at lower values of $\ell$ as the phase transition is approached, reflecting the behavior of the homogeneous condensate \cite{Abad2013}. For high enough $\Omega_R$ some modes (typically with spin character) may become energetically unstable, and the persistent currents decay.

\begin{figure}\centering
	\includegraphics[clip=true,width=\linewidth]{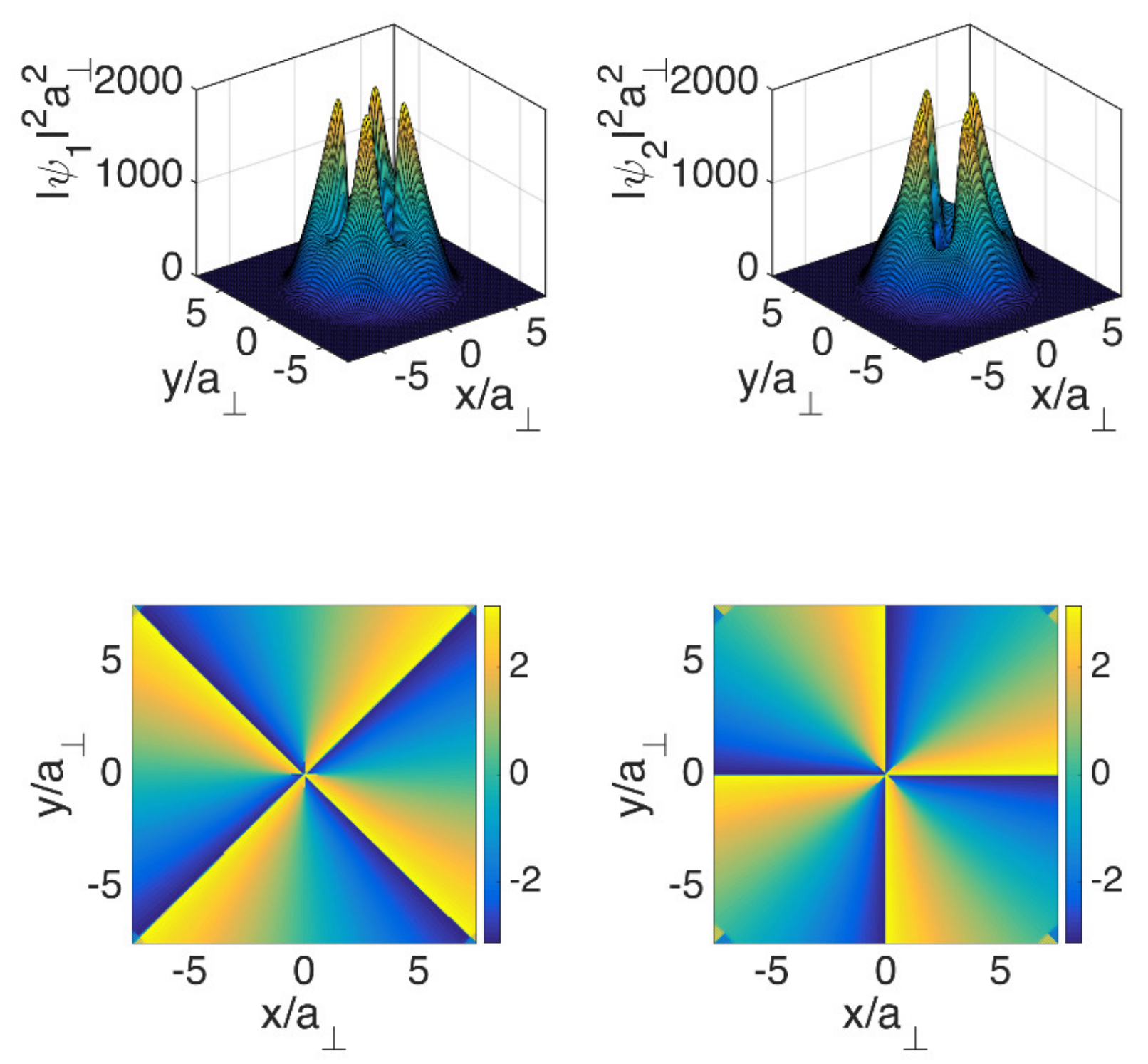}
	\caption{Density (top panels) and phase (bottom panels) of a symmetry broken configuration with $\kappa=4$, $\Omega_R=1.7\,\hbar\omega_\perp$ and $g_{12}=1.5g$.}\label{Fig10}
\end{figure}

For $\Omega_R$ above the highly unstable region, the spin mode drives a symmetry breaking that breaks the rotational symmetry and creates inhomogeneous densities along the azimuthal direction. 
These density structures can be understood as follows. When $\Omega_R$ decreases towards the phase transition from the GS1 side, there is an energetic instability of the persistent currents with centered singularities (vortices), as implied in Fig.~\ref{Fig8} for $\Omega_R<1.95\,\hbar\omega_\perp$. 
The system then undergoes a symmetry breaking of the phase distribution and starts the decay process by expelling the vortices in both components away from the center. As soon as the vortices have started moving, they induce a change in the density which is magnified by the large fluctuations close to the critical point. The azimuthal dependence of the density induces a redistribution of energy contributions that allows the two-component BEC to keep the currents, and guarantees a zero global polarization even below the phase transition. In this new situation, the system becomes both energetically and dynamically stable, with all eigenfrequencies positive and real (shaded area in Fig.~\ref{Fig8}), although they may be very small.  We have checked the stablity of these structures by evolving them in real-time simulations of Eqs.~\eqref{Eq:GP1} and \eqref{Eq:GP2}, and finding no appreciable change. 
An example of a symmetry broken configuration is shown in Fig.~\ref{Fig10}, where we plot the density distributions for $\Omega_R=1.7\,\hbar\omega_\perp$ and $\kappa=4$. Four clear out-of-phase density peaks are formed in both components, reflecting the spin-density character. We can see from the phase plots that the vortices causing the persistent currents are moved slightly from the origin.

\section{Discussion and Conclusions}\label{SecConcl}

We have studied the mean-field theory of persistent currents in a coherently coupled two-component condensate in a ring trap, both in the neutral and the polarized phases (in our notation, GS1 and GS2). In the neutral phase we have seen that by modifying the intensity of the lasers generating the coupling (that is the Rabi frequency), or equivalently by modifying the ratio of inter- and intra-component scattering lengths (via, for instance, a Feshbach resonance), the decay of the persistent currents is related to an energetic instability of the density or the spin-density modes. 

When the decay in GS1 is driven by the density mode, diagonalization of the full Bogoliubov operator shows that close to instability there appears a roton-like structure that is linked to the nucleation of excitations in the internal surface, similarly to what has been predicted in single components \cite{Dubessy2012} and spinor condensates~\cite{Makela2013}. This roton-like structure induces a minimum in the Doppler-shifted excitation spectrum at finite $\ell$, which leads to the decay of the persistent current through phase-slips (vortices) when the lowest frequency becomes negative. Instead, when the decay in GS1 is driven by the spin-density mode, the finite-$\ell$ energetic instability emerges naturally due to the presence of the gap in the excitation spectrum and no roton-like minimum can be seen.

Roton minima in the excitation spectrum are very well known in helium superfluids, and are an effect of the many-body character of the interactions. In BECs they have been predicted not only in static situations such as in dipolar condensates~\cite{Santos2003}, but also in the presence of currents in the ground state, such as in spin-orbit coupled condensates~\cite{Zheng2012,Martone2012} or the results presented here and in Refs.~\cite{Dubessy2012,Makela2013}. 
In the dipolar case, the roton mode touching zero is generally associated to density-wave or supersolid configurations (see the recent experiment \cite{Kadau2015}). Instead, in the case presented in this article, the roton mode touching zero drives an energetic instability that leads to the nucleation of vortices in the internal surface of the ring and the subsequent decay of the persistent currents. The finite quasimomentum associated with the roton is a necessary ingredient to allow the process of vortex nucleation which causes the superflow decay. The absence of the roton in the non-interacting limit reflects the fact that in a non-interacting Bose gas, described by the (linear) Schr\"odinger equation, symmetry breakings do not take place and all states of finite winding number are (meta)stable states. 

In the polarized phase the conditions for stability and decay of persistent currents are more complicated. At large values of $\Omega_R$ an unstable spin-density mode can drive a symmetry breaking in the densities of the two components that allows the system to preserve the zero polarization and keep the currents. At smaller values of $\Omega_R$, the system shows a finite polarization and energetic instabilities drive the decay of the currents. The character of these unstable modes is not purely density or spin, but may be hybridized or, typically for very large polarizations, may acquire the character of minority or majority excitation modes. In all the cases, however, the final state of the decay process gives equal angular momentum to both components.

Notice that the situation we have considered studies the system close to its ground state or stationary metastable states, that is the phase difference between components is always locked to $\pi$ (except for small fluctuations) and no counterflow is allowed. Relaxation of these conditions leads to very interesting soliton and vortex dynamics~\cite{Usui2015,Gallemi2015}, closely related to the special kind of soliton introduced in \cite{Son2002}.

It would be very interesting to test the results found in this work experimentally, especially what concerns the different nature of the density and spin excitations and the symmetry breaking driven by the closeness to the transition to the GS2 phase. In particular, the latter would be an indirect proof of the fact that it is vortices (hidden in the inner empty region) that create the persistent currents. The different technologies required are available: condensates in ring traps are made in laboratories, coherenty coupled condensates have been experimentally studied, and both Feshbach resonances and Rabi couplings have been used. In Ref.~\cite{Beattie2013} persistent currents with $\kappa=3$ were shown to be stable in a two-component $^{87}$Rb mixture when the Rabi coupling was kept during the time evolution, reflecting the fact that the opening of the gap in the spin mode in the presence of $\Omega_R$ drove the system away from the spin instability. By studying a higher winding number and different values of $\Omega_R$, the GS1 decay dynamics could be tested. 
Recent experiments and new techniques on the measurement of phonon modes in persistent currents~\cite{Edward2015,Kumar2015} also offer new perspectives, and might allow an observation of the current-induced roton mode.

\acknowledgments

I thank Thomas Busch, Rashi Sachdeva, Muntsa Guilleumas, Ricardo Mayol, Antonio Mu\~noz and Albert Gallem\'i for a careful reading of the mansucript. Discussions with Tadhg Morgan and Angela White are also acknowledged. This work has been supported by the Okinawa Institute of Science and Technology Graduate University.

\end{document}